%% file: main.tex
\documentclass[ALICE,manyauthors]{cernphprep}
\usepackage[comma,square,numbers,sort&compress]{natbib}
\usepackage{lineno}
\usepackage{xspace}
\usepackage{hyperref}
\usepackage{comment}
\usepackage{rotating}
\usepackage[utf8]{inputenc}
\usepackage{amsmath}

\usepackage[T1]{fontenc}
\usepackage{orcidlink}

\begin{document}
\input{commands.tex}

\begin{titlepage}
\PHyear{2024}       
\PHnumber{302}      
\PHdate{12 Nov}  

\title{Studying charm hadronisation into baryons with azimuthal correlations of $\Lambda_{\rm c}^{+}$ with charged particles in pp collisions at $\mathbf{\sqrt{\textit s} = 13}$ TeV}
\ShortTitle{Azimuthal correlations of $\Lambda_{\rm c}^{+}$ with charged particles in pp collisions}   

\Collaboration{ALICE Collaboration\thanks{See Appendix~\ref{app:collab} for the list of collaboration members}}
\ShortAuthor{ALICE Collaboration} 

\begin{abstract}
The distribution of angular correlations between prompt charm hadrons and primary charged particles in pp collisions is sensitive to the charm-quark hadronisation process. In this letter, charm-baryon correlations are measured for the first time by studying the azimuthal-angle difference between charged particles and prompt $\lc$ baryons produced in pp collisions at a centre-of-mass energy of $\s = 13$ \TeV, with the ALICE detector. 
$\Lc$ baryons are reconstructed at midrapidity ($|y| < 0.5$) in the transverse-momentum interval $3 < \pt < 16$ \GeVc, and correlated with charged particles with $\pt > 0.3$ \GeVc and pseudorapidity $|\eta| < 0.8$.
For $3<\ptLd<5,$~\GeVc, the comparison with published measurements of D-meson and charged-particle correlations in the same collision system hints at a larger number of low-momentum particles associated with $\Lc$-baryon triggers than with D-meson triggers, both in the collinear and opposite directions with respect to the trigger particle.
These differences can be quantified by the comparison of the properties of the near- and away-side correlation peaks, and are not reproduced by predictions of various Monte Carlo event generators, generally underpredicting the associated particle yields at $\ptass<1$~\GeVc. This tension between $\Lc$-baryon and D-meson associated peak yields could suggest a modified fragmentation of the charm quark, or a different hadronisation process, when a charm baryon is produced in the final state.
\end{abstract}
\end{titlepage}

\setcounter{page}{2} 


\input{Introduction} 
\input{ExpermentalSetUp}
\input{AnalysisStrategy}
\input{Systematics}
\input{Results}

\input{Conclusions}


\newenvironment{acknowledgement}{\relax}{\relax}
\begin{acknowledgement}
\section*{Acknowledgements}
\input{fa_2024-11-02_Opt_C.tex}
\end{acknowledgement}

\bibliographystyle{utphys}   
\bibliography{bibliography}

\newpage
\appendix

%
%

\section{The ALICE Collaboration}
\label{app:collab}
\input{Alice_Authorlist_2024-11-02_Opt_C.tex}
\end{document}

%% file: commands.tex
%

\newcommand{\pp}           {pp\xspace}
\newcommand{\ppbar}        {\mbox{$\mathrm {p\overline{p}}$}\xspace}
\newcommand{\XeXe}         {\mbox{Xe--Xe}\xspace}
\newcommand{\PbPb}         {\mbox{Pb--Pb}\xspace}
\newcommand{\pA}           {\mbox{pA}\xspace}
\newcommand{\pPb}          {\mbox{p--Pb}\xspace}
\newcommand{\AuAu}         {\mbox{Au--Au}\xspace}
\newcommand{\dAu}          {\mbox{d--Au}\xspace}

\newcommand{\s}            {\ensuremath{\sqrt{s}}\xspace}
\newcommand{\snn}          {\ensuremath{\sqrt{s_{\mathrm{NN}}}}\xspace}
\newcommand{\pt}           {\ensuremath{p_{\rm T}}\xspace}
\newcommand{\meanpt}       {$\langle p_{\mathrm{T}}\rangle$\xspace}
\newcommand{\ycms}         {\ensuremath{y_{\rm CMS}}\xspace}
\newcommand{\ylab}         {\ensuremath{y_{\rm lab}}\xspace}
\newcommand{\etarange}[1]  {\mbox{$\left | \eta \right |~<~#1$}}
\newcommand{\yrange}[1]    {\mbox{$\left | y \right |~<~#1$}}
\newcommand{\dndy}         {\ensuremath{\mathrm{d}N_\mathrm{ch}/\mathrm{d}y}\xspace}
\newcommand{\dndeta}       {\ensuremath{\mathrm{d}N_\mathrm{ch}/\mathrm{d}\eta}\xspace}
\newcommand{\avdndeta}     {\ensuremath{\langle\dndeta\rangle}\xspace}
\newcommand{\dNdy}         {\ensuremath{\mathrm{d}N_\mathrm{ch}/\mathrm{d}y}\xspace}
\newcommand{\Npart}        {\ensuremath{N_\mathrm{part}}\xspace}
\newcommand{\Ncoll}        {\ensuremath{N_\mathrm{coll}}\xspace}
\newcommand{\dEdx}         {\ensuremath{\textrm{d}E/\textrm{d}x}\xspace}
\newcommand{\RpPb}         {\ensuremath{R_{\rm pPb}}\xspace}

\newcommand{\nineH}        {$\sqrt{s}~=~0.9$~Te\kern-.1emV\xspace}
\newcommand{\seven}        {$\sqrt{s}~=~7$~Te\kern-.1emV\xspace}
\newcommand{\twoH}         {$\sqrt{s}~=~0.2$~Te\kern-.1emV\xspace}
\newcommand{\twosevensix}  {$\sqrt{s}~=~2.76$~Te\kern-.1emV\xspace}
\newcommand{\five}         {$\sqrt{s}~=~5.02$~Te\kern-.1emV\xspace}
\newcommand{\twosevensixnn}{$\sqrt{s_{\mathrm{NN}}}~=~2.76$~Te\kern-.1emV\xspace}
\newcommand{\fivenn}       {$\sqrt{s_{\mathrm{NN}}}~=~5.02$~Te\kern-.1emV\xspace}
\newcommand{\LT}           {L{\'e}vy-Tsallis\xspace}
\newcommand{\GeVc}         {Ge\kern-.1emV/$c$\xspace}
\newcommand{\MeVc}         {Me\kern-.1emV/$c$\xspace}
\newcommand{\TeV}          {Te\kern-.1emV\xspace}
\newcommand{\GeV}          {Ge\kern-.1emV\xspace}
\newcommand{\MeV}          {Me\kern-.1emV\xspace}
\newcommand{\tev}          {Te\kern-.1emV\xspace}
\newcommand{\gev}          {\rm Ge\kern-.1emV\xspace}
\newcommand{\mev}          {\rm Me\kern-.1emV\xspace}
\newcommand{\GeVmass}      {Ge\kern-.2emV/$c^2$\xspace}
\newcommand{\MeVmass}      {Me\kern-.2emV/$c^2$\xspace}
\newcommand{\lumi}         {\ensuremath{\mathcal{L}}\xspace}

\newcommand{\ITS}          {\rm{ITS}\xspace}
\newcommand{\TOF}          {\rm{TOF}\xspace}
\newcommand{\ZDC}          {\rm{ZDC}\xspace}
\newcommand{\ZDCs}         {\rm{ZDCs}\xspace}
\newcommand{\ZNA}          {\rm{ZNA}\xspace}
\newcommand{\ZNC}          {\rm{ZNC}\xspace}
\newcommand{\SPD}          {\rm{SPD}\xspace}
\newcommand{\SDD}          {\rm{SDD}\xspace}
\newcommand{\SSD}          {\rm{SSD}\xspace}
\newcommand{\TPC}          {\rm{TPC}\xspace}
\newcommand{\TRD}          {\rm{TRD}\xspace}
\newcommand{\VZERO}        {\rm{V0}\xspace}
\newcommand{\VZEROA}       {\rm{V0A}\xspace}
\newcommand{\VZEROC}       {\rm{V0C}\xspace}
\newcommand{\Vdecay} 	   {\ensuremath{V^{0}}\xspace}

\newcommand{\ee}           {\ensuremath{e^{+}e^{-}}} 
\newcommand{\pip}          {\ensuremath{\pi^{+}}\xspace}
\newcommand{\pim}          {\ensuremath{\pi^{-}}\xspace}
\newcommand{\kap}          {\ensuremath{\rm{K}^{+}}\xspace}
\newcommand{\kam}          {\ensuremath{\rm{K}^{-}}\xspace}
\newcommand{\pbar}         {\ensuremath{\rm\overline{p}}\xspace}
\newcommand{\kzero}        {\ensuremath{{\rm K}^{0}_{\rm{S}}}\xspace}
\newcommand{\lmb}          {\ensuremath{\Lambda}\xspace}
\newcommand{\almb}         {\ensuremath{\overline{\Lambda}}\xspace}
\newcommand{\Om}           {\ensuremath{\Omega^-}\xspace}
\newcommand{\Mo}           {\ensuremath{\overline{\Omega}^+}\xspace}
\newcommand{\X}            {\ensuremath{\Xi^-}\xspace}
\newcommand{\Ix}           {\ensuremath{\overline{\Xi}^+}\xspace}
\newcommand{\Xis}          {\ensuremath{\Xi^{\pm}}\xspace}
\newcommand{\Oms}          {\ensuremath{\Omega^{\pm}}\xspace}
\newcommand{\degree}       {\ensuremath{^{\rm o}}\xspace}

\newcommand{\Dzero}     {\ensuremath{\rm D}^{0}\xspace}
\newcommand{\Dplus}     {\ensuremath{\rm D}^{+}\xspace}
\newcommand{\Dstar}     {\ensuremath{\rm D}^{*+}\xspace}
\newcommand{\Dstarwide} {\ensuremath{\rm D}^{*}(2010)^{+}\xspace}
\newcommand{\Lc}        {\ensuremath{\rm \Lambda}_{\rm c}^{+}\xspace}
\newcommand{\lc}        {\ensuremath{\rm \Lambda}_{\rm c}^{+}\xspace}
\newcommand{\Sc}        {\ensuremath{\rm \Sigma}_{\rm c}^{0,++}\xspace}
\newcommand{\Dphi}      {\Delta\varphi}
\newcommand{\dphi}      {\Delta\varphi}
\newcommand{\Deta}      {\Delta\eta}
\newcommand{\deta}      {\Delta\eta}
\newcommand{\ptD}       {\ensuremath{p_{\rm T}}^{\rm D}\xspace}
\newcommand{\ptDL}      {\ensuremath{p_{\rm T}}^{\rm D,\,\Lambda_{\rm c}^{+}}\xspace}
\newcommand{\ptL}       {\ensuremath{p_{\rm T}}^{\rm \Lambda_{\rm c}^{+}}\xspace}
\newcommand{\ptl}       {\ensuremath{p_{\rm T}}^{\rm \Lambda_{\rm c}^{+}}\xspace}
\newcommand{\ptLd}       {\ensuremath{p_{\rm T}}^{\rm \Lambda_{\rm c}^{+},\rm{D}}\xspace}
\newcommand{\ptass}     {\ensuremath{p_{\rm T}}^{\rm assoc}\xspace}

%% file: Introduction.tex
\section{Introduction}
\label{sec:Intro}

In recent years, the investigation of charm-quark hadronisation has emerged as a pivotal field of research within high-energy particle physics.
Following a factorisation approach based on the separation of processes involving
soft and hard squared momentum-transfer ($Q^2$) scales~\cite{Collins:1989gx}, the production cross section of charm hadrons in ultrarelativistic hadronic collisions can be described as the convolution of three terms: (i) the parton distribution functions (PDFs) of the colliding nucleons; (ii) the cross section of the hard parton scattering responsible for generating the ${\rm c}\overline{{\rm c}}$ pair, which can be calculated using perturbative quantum chromodynamics (pQCD) due to the large $Q^2$ of the process; and (iii) the charm-quark fragmentation function, governing the transition from the charm quark into a final-state hadron. The fragmentation function cannot be computed perturbatively, as it is associated with soft processes featuring large values of the QCD coupling constant $\alpha_{\rm s}$. Consequently, it is generally parameterised from measurements performed in ${\rm e}^+{\rm e}^-$ and ep collisions, assuming its universality across different collision systems~\cite{ParticleDataGroup:2024cfk}.

While measurements of D-meson production at hadronic colliders are well described by pQCD calculations employing fragmentation fractions extracted from ${\rm e}^+{\rm e}^-$ and ep collisions~\cite{ALICE:2019nxm,ALICE:2021npz,ALICE:2023sgl,ATLAS:2015igt,CMS:2021lab,LHCb:2016ikn,LHCb:2015swx}, in the last decade a large wealth of results related to charm-baryon production have disproved the fragmentation fraction universality assumption.
In particular, measurements of $\Lc/\Dzero$ production yield ratios in pp collisions at the LHC showed an enhancement for \pt $<$ 8 \GeVc compared to equivalent measurements in ${\rm e}^+{\rm e}^-$ and ep collisions, as well as to expectations from pQCD calculations and Monte Carlo simulations using a model of quark fragmentation constrained by measurements on ${\rm e}^+{\rm e}^-$ and ep collisions~\cite{ALICE:2022exq,ALICE:2020wla,ALICE:2020wfu,ALICE:2021rzj,CMS:2019uws}. An even larger enhancement was observed for the production of charm-strange baryon states compared to $\Dzero$ production~\cite{ALICE:2021psx,ALICE:2021bli,ALICE:2022cop}.
As a consequence, the fragmentation fractions of charm quarks into ground-state charm hadrons measured in pp collisions at the LHC~\cite{ALICE:2021dhb, ALICE:2023sgl} show significantly larger values for charm baryons and, consequently, reduced values for D mesons, with respect to those measured in ${\rm e}^+{\rm e}^-$ and ep collisions~\cite{Lisovyi:2015uqa,Azizi:2014nta}. 

Several models try to explain the observed baryon enhancement in terms of modified hadronisation with respect to in-vacuum fragmentation, either by considering colour reconnection mechanisms beyond the leading-colour approximation, with new junction topologies that favour baryon formation~\cite{Christiansen:2015yqa}, or by considering coalescence as an accompanying hadronisation mechanism~\cite{Song:2018tpv, Minissale:2020bif}. Another approach (``SHM+RQM"~\cite{He:2019tik}) foresees the existence of a significant number of unobserved charm-baryon resonant states, whose decay would contribute to the charm-baryon ground-state production yields.
While these models can generally reproduce the $\Lc/\Dzero$ measurements, they have more difficulties predicting the enhancement of higher-mass baryons as $\Xi_{\rm c}^{0,+}$ and $\Omega_{\rm c}^{0}$~\cite{ALICE:2023sgl,ALICE:2022cop}. A comprehensive understanding of charm hadronisation in hadronic collisions thus remains challenging.

On top of baryon-to-meson particle ratios, additional information on possible modifications of charm hadronisation was obtained from a study of the jet momentum fraction carried by the $\Lc$ baryon along the jet axis~\cite{ALICE:2023jgm}, and from a comparison with analogous results for D-meson jets~\cite{ALICE:2022mur}, performed with ALICE in pp collisions at $\sqrt{s}$ = 13 TeV.
Studying jets containing a $\Lc$ baryon and their kinematic structure also allows us to probe whether charm hadronisation features are related to a local effect, i.e. related to particles close in phase-space to the $\Lc$ baryon.
The results obtained from the measurement~\cite{ALICE:2023jgm} hint at a softer fragmentation of the charm quark when it fragments into a baryon, at least within the limited kinematic region probed.

Further investigation on the charm hadronisation mechanism and the charm-jet properties can be obtained from specific studies of angular correlations involving charm hadrons.
In particular, for the ${\rm c}\overline{{\rm c}}$ pair production process at leading-order in $\alpha_{\rm s}$, the azimuthal correlation distribution between charm hadrons (denoted as ``trigger particles'') and charged (``associated'') particles produced in the same collision features a ``near-side'' peak at $\Dphi=\varphi_{\mathrm{trigger}}-\varphi_{\mathrm{assoc}} \approx 0$ and an ``away-side'' peak at $\Dphi \approx \pi$. The former peak arises from the particles produced by the fragmentation of the charm quark that hadronises into the trigger charm hadron, while the latter is originated by the other charm-quark fragmentation products~\cite{ALICE:2021kpy}.
The shape, height and integral of the near-side peak provide a detailed characterisation of the charm-induced jet. Such properties are closely related to the amount of charged particles produced from the charm parton shower in association with the charm hadron, their angular displacement from such a hadron, and their transverse momentum. All these quantities are, in turn, directly influenced by the specific mechanism of hadronisation of the charm quark.
A complete characterisation of the near-side peak of the correlation distribution in the case of charm-quark hadronisation into a D meson has already been performed by ALICE~\cite{ALICE:2016clc,ALICE:2019oyn,ALICE:2021kpy}, allowing us to evaluate the properties of charm jets and to validate predictions from Monte Carlo generators and models. 
For charm baryons, no corresponding results are currently available.

In this Letter, the first measurements of the azimuthal correlation distribution between prompt $\Lc$ baryons and charged particles in pp collisions are reported. Prompt $\Lc$ baryons are defined as those produced directly in the hadronisation of charm quarks or originating from the decays of directly-produced excited charm-hadron states, hence excluding weak decays of beauty hadrons. The study uses a sample of pp collisions at $\sqrt{s}$ = 13 TeV, collected with the ALICE detector during the LHC Run~2. A quantitative estimation of the correlation peak features is provided in different transverse-momentum ranges of both $\Lc$ baryons and charged particles, focusing in particular on the near-side peak. The results are compared to previous ALICE measurements of D-meson azimuthal correlations with charged particles to explore potential differences in the charm hadronization process associated with baryon formation. A comparison of the results to predictions from Monte Carlo generators considering either in-vacuum fragmentation or modified hadronisation mechanisms is also reported.

The Letter is structured as follows. Section~\ref{sec:ALICEdetector} summarises information on the ALICE detector, data and Monte Carlo sample exploited for the study. The analysis procedure is described in Sec.~\ref{sec:analysis}. The systematic uncertainties affecting the measurement are outlined in Sec.~\ref{sec:Systematics}. A discussion of the results obtained from the measurement is reported in Sec.~\ref{sec:Results}, and conclusions are provided in Sec.~\ref{sec:Conclusions}.

%% file: ExpermentalSetUp.tex
\section{Experimental apparatus and data samples}
\label{sec:ALICEdetector}

A comprehensive description of the ALICE detector and its operational characteristics can be found in Refs.\cite{Aamodt:2008zz, Abelev:2014ffa}. The reconstruction of $\Lc$ baryons and charged particles was carried out using detectors located in the central barrel, covering a pseudorapidity of $|\eta| < 0.9$ and subject to a magnetic field of 0.5 T aligned parallel to the beam axis. In particular, the Inner Tracking System (ITS)~\cite{Aamodt:2010aa} and the Time Projection Chamber (TPC)~\cite{Alme:2010ke} were used to reconstruct the tracks of charged particles. The primary interaction vertex and the decay vertices of charm hadrons were reconstructed by exploiting the excellent spatial resolution provided by the ITS.
The TPC, in conjunction with the Time-of-Flight (TOF) detector~\cite{Akindinov:2013tea}, provided information for charged-particle identification (PID). Additionally, detectors positioned along the beamline, covering forward and backward rapidity, played a crucial role in the analysis. The V0 detector~\cite{Abbas:2013taa} is a set of scintillators covering the pseudorapidity ranges \mbox{$2.8 < \eta < 5.1$} (V0A) and \mbox{$-3.7 < \eta <-1.7$} (V0C), used for event triggering and background rejection. The T0 detector is an array of Cherenkov counters, positioned along the beamline, at a distance of $+370$ cm (T0A) and $-70$ cm (T0C) from the nominal interaction point, which provided the collision starting time used by the TOF~\cite{Adam:2016ilk}.

The data sample employed in the analysis comprised proton--proton (pp) collisions at $\s = 13$ TeV, recorded in 2016, 2017, and 2018, and consisting of an integrated luminosity of $\mathcal{L}_{\rm int} = 29.2 \pm 0.5~{\rm nb}^{-1}$, determined from the visible cross section measured with the V0 detector~\cite{ALICE-PUBLIC-2016-002}. 
The collisions were recorded using a minimum bias (MB) trigger, requiring coincident signals in both the V0 scintillators.
Background events arising from interactions between protons in the beam and residual gas within the beam pipe were removed in the offline processing using the time information from the V0 and the correlation between the number of hits and the track segments reconstructed in the Silicon Pixel Detector (SPD), which constitutes the two innermost layers of the ITS. Furthermore, all events featuring more than one reconstructed primary vertex with at least five reconstructed tracks associated were eliminated in order to exclude pileup events within the same bunch crossing.
Only events with a reconstructed primary vertex within $\pm 10$ cm from the nominal centre of the ALICE detector along the beam direction were selected in order to ensure a uniform acceptance of the central-barrel detectors.

Monte Carlo (MC) samples of pp collisions at the same centre-of-mass energy were employed for training the machine-learning algorithm used to discriminate between signal and background $\Lc$ baryon candidates, as well as for correcting the azimuthal-correlation measurements.
The Monte Carlo samples were produced with the PYTHIA~8.243 event generator~\cite{Sjostrand:2014zea}, with SoftQCD and using the Monash tune~\cite{Skands:2014pea}, and requiring that each collision contained either a ${\rm c}\overline{\rm c}$ or a ${\rm b}\overline{\rm b}$ pair.
These simulations include the transport of the produced particles through the ALICE detector via the GEANT3 package~\cite{Brun:1082634}, reproducing its complete geometry, response, and conditions throughout the data acquisition.

%% file: AnalysisStrategy.tex
\section{Analysis strategy}
\label{sec:analysis}

The analysis strategy follows the procedure outlined in previous studies on angular correlations of D mesons with primary charged particles~\cite{ALICE:2016clc, ALICE:2019oyn, ALICE:2021kpy}. The main analysis steps include: (i) selecting $\lc$ and charged particles; (ii) evaluating the azimuthal-correlation distribution and applying corrections for detector-related effects, contamination from secondary particles, and contribution from beauty feed-down; (iii) fitting the azimuthal-correlation distribution of $\lc$ baryons with charged particles to extract quantitative information on its properties.

\subsection{Selection of \texorpdfstring{$\lc$}{Lc} baryons and associated particles}

$\lc$ baryons, reconstructed from the hadronic decay channel $\lc \to \rm p K^{-}\pi^{+}$ in the transverse-momentum interval 3 $< \ptL < $ 16 GeV/$c$, were used, together with their charge conjugates, as trigger particles in this analysis.
The total branching ratio of this channel, considering both resonant and direct decay channels, is B.R.~=~$(6.24 \pm 0.28)\%$~\cite{ParticleDataGroup:2024cfk}. A binary classification approach based on Boosted Decision Trees (BDT) algorithms provided by the XGBOOST library~\cite{2016arXiv160302754C} was used to optimally suppress the combinatorial background, enhancing the statistical significance of the signal invariant-mass peak, while keeping the signal efficiency as high as possible~\cite{ALICE:2021rzj}. The algorithm considered topological variables which exploit the characteristic displacement of the $\lc$ decay vertex with respect to the primary vertex (c$\tau~\approx~60~\mu \rm{m}$~\cite{ParticleDataGroup:2024cfk}). The particle-identification information of TPC and TOF on the $\lc$ decay products were also employed.
The $\lc$ raw yields were extracted with a fit to the invariant-mass distribution of the selected candidates, performed using a second-order polynomial to describe the background and a Gaussian function to model the signal peak, as in similar analyses~\cite{ALICE:2021npz}. Such a choice allows for having good quality fits, with $\chi^{2}/$ndf values lying in the range $0.8-1.4$ depending on the $\lc$ transverse-momentum range.

Associated particles were defined as primary charged particles~\cite{ALICE-PUBLIC-2017-005} (being pions, kaons, protons, electrons, or muons) with transverse momentum $\ptass>0.3$~\GeVc and pseudorapidity $|\eta| <$ 0.8. The decay products of the $\lc$ candidates were excluded from the associated-particle sample.~Charged-particle tracks were reconstructed by requiring a minimum of 70 space points out of 159 in the TPC, 2 out of 6 in the ITS, and a $\chi^2$/ndf of the momentum fit in the TPC smaller than 2. Furthermore, tracks were required to have a maximum distance of closest approach (DCA) to the primary vertex of 1 cm in the transverse ($xy$) plane and along the beamline ($z$ direction).

\subsection{Construction of the azimuthal-correlation distribution and corrections}

The selected $\lc$ baryon candidates within $\pm 2\sigma$ from the centre of the invariant-mass signal peak (referred to as the ``peak region") were correlated with associated particles reconstructed and selected within the same event (SE). This choice was preferred to the more common $3\sigma$ definition to reduce the contribution of the combinatorial background in the peak region. A two-dimensional angular-correlation distribution, denoted as $C_{\rm SE}(\Delta\varphi,\Delta\eta)_{\rm{peak~region}}$, was built for each kinematic range studied. 
The resulting angular correlation distribution was corrected for the limited detector acceptance and efficiency for the reconstruction and selection of $\lc$ baryon candidates and associated particles. This was done by weighting each correlation pair by $1/(A\times\epsilon)^{\rm{assoc}}$ $\times$ $1/(A\times\epsilon)^{\rm{trigger}}$, where $A$ and $\epsilon$ represent the acceptance and efficiency factors of the associated particles and the $\lc$ baryons, respectively, evaluated using Monte Carlo simulations and assumed to be independent of each other.
The contribution of background $\lc$ candidates to the correlation distribution was estimated using the two-dimensional angular-correlation distribution evaluated in the $\lc$ sidebands, $C_{\rm SE}(\Delta\varphi,\Delta\eta)_{\rm{sidebands}}$, considering the $\lc$ candidates with invariant mass $M$ in the range $4\sigma < |M-\mu| < 8\sigma$ as triggers, with $\mu$ being the centre of the invariant-mass signal peak. 
This contribution was normalised by a factor $B_{\rm peak} \left/  B_{\rm sidebands} \right.$, where $B_\mathrm{peak}$ and $B_\mathrm{sidebands}$
denote the number of $\lc$ background candidates in the peak and sideband regions, respectively. These were evaluated as the integral of the background component of the fit function to the invariant-mass distribution in those regions. The subtraction from the signal region correlation distribution of the normalised sideband distribution was then performed.

A correction for losses due to pair acceptance effects and detector inhomogeneities was estimated using the mixed-event~(ME) technique. Specifically, mixed-event correlation distributions were built in the peak region and the sidebands ($C_{\rm ME}(\Delta\varphi,\Delta\eta)_{\rm{peak~region}}$ and $C_{\rm ME}(\Delta\varphi,\Delta\eta)_{\rm{sidebands}}$, respectively) by correlating $\lc$-trigger candidates with associated charged particles from other events characterised by a similar midrapidity event multiplicity and primary vertex location along the beam axis, and applying the same acceptance and efficiency weights used for the same-event distributions. 
After normalising the $C_{\rm ME}(\Delta\varphi,\Delta\eta)_{\rm{peak~region}}$ and $C_{\rm ME}(\Delta\varphi,\Delta\eta)_{\rm{sidebands}}$ distributions to 1 in the range $(\Delta\varphi,\Delta\eta) \approx (0,0)$, the mixed-event correction was applied by dividing the same-event correlation distributions by the mixed-event ones. 

To ensure adequate statistical precision for this analysis, the per-trigger normalised two-dimensional correlation distribution was integrated within the range $|\Delta\eta| < 1$, and the correlation distribution was restricted to the interval $0 \leq \Delta\varphi \leq \pi$ by reflecting the original correlation distribution, which covered the range $-\pi/2 \leq \Delta\varphi \leq 3\pi/2$, around the symmetry points $\Delta\varphi= 0$ and $\Delta\varphi=\pi$. 

A correction factor for removing the residual contamination of non-primary associated particles not rejected by the track selection, $p_\mathrm{prim}(\Dphi)$, was applied differentially in $\Dphi$. It was evaluated with Monte Carlo simulations based on PYTHIA~8 by quantifying the fraction of primary particles, among all the reconstructed tracks, satisfying the selection criteria, as detailed in~\cite{ALICE:2019oyn}. 
The above procedure is summarised in the following equation:

\begin{equation}
\label{eq:Corr}
\centering
    \Tilde{C}_{\rm{inclusive}}(\Dphi, \Delta\eta) = \frac{p_\mathrm{prim}(\Dphi)}{S_\mathrm{peak}}\left( \frac{C_{\mathrm{\,SE}}(\Dphi, \Delta\eta)}{C_\mathrm{\,ME}(\Dphi, \Delta\eta)} \bigg|_\mathrm{peak~region} - \frac{B_\mathrm{peak}}{B_\mathrm{sidebands}}\frac{C_\mathrm{\,SE}(\Dphi, \Delta\eta)}{C_\mathrm{\,ME}(\Dphi, \Delta\eta)} \bigg|_\mathrm{sidebands}\right), 
\end{equation}

where $S_\mathrm{peak}$ is the number of triggers in the peak region, evaluated by integrating in the peak region the Gaussian component of the fit function applied to the invariant-mass distribution, weighted by $1/(A\times\epsilon)^{\rm{trigger}}$. It provides the per-trigger normalisation of the signal correlation distribution.

A large fraction $f_{\rm{prompt}}$ of the reconstructed $\lc$ baryons originate from charm-quark hadronisation or decays of excited open charm or quarkonium states. The value of $f_{\rm{prompt}}$ ranges from 92\% to 85\%, decreasing with increasing $\ptL$. The remaining $\lc$ baryons, known as feed-down $\lc$ baryons, are produced from decays of beauty hadrons. The feed-down contribution to the measured correlation distribution was estimated using templates of the azimuthal-correlation distribution of feed-down $\lc$ baryons with charged particles, $\Tilde{C}^\mathrm{MC}_\mathrm{feed-down}(\Dphi)$, obtained with PYTHIA~8 simulations\footnote{For data-to-model comparisons and for getting correction factors that did not require a simulation of the detector response, PYTHIA 8.304 was used. Hereafter, for brevity, such a version will be referred to as PYTHIA~8.}.

The topological selections applied to the non-prompt $\lc$-baryon candidates favour specific beauty-hadron decay topologies, where the $\lc$-baryon and the other decay charged particles are emitted in similar directions. As a consequence, the shape of the feed-down correlation contribution present in the data distribution was affected by distortions at small $\Delta\varphi$ values compared to the template function evaluated from the PYTHIA~8 simulations.
A correction factor $b_{\rm{bias}}(\Delta\varphi)$, obtained via Monte Carlo studies, was applied to the measured correlation distribution $\Tilde{C}_\mathrm{inclusive}(\Delta\varphi)$ to account for such effect, as detailed in Ref.~\cite{ALICE:2019oyn}. The feed-down correlation template was then subtracted from the measured distribution after normalising it by ($1 - f_{\rm{prompt}}$), i.e.~the fraction of feed-down $\lc$ candidates, evaluated exploiting FONLL calculations for the beauty-quark production cross section~\cite{Cacciari:1998it,Cacciari:2001td,Cacciari:2012ny} and LHCb parton-to-hadron fragmentation fractions~\cite{LHCb:2019fns}. 

As recently measured by ALICE~\cite{ALICE:2021rzj}, a significant fraction (approximately $\approx 38\%$) of $\lc$ baryons originate from the decay of $\Sigma_\mathrm{c}^{0,+,++} (2455)$ baryons. Among these states, the $\Sigma_\mathrm{c}^{0,++} (2455)$ decays into a $\lc$ baryon paired with a low-momentum pion, with a branching ratio of 100$\%$~\cite{ParticleDataGroup:2024cfk}. These low-momentum pions tend to be predominantly aligned with the direction of the $\lc$ baryon, resulting in an enhancement of the azimuthal correlation distribution around $\Delta\varphi=0$. This contribution, not related to the charm parton shower and hadronisation, was removed by simulating the $\Sigma_\mathrm{c}^{0,++}\to\lc\pi^{-, +}$ decay kinematics and computing the azimuthal correlation distribution between their decay products ($\Tilde{C}^\mathrm{MC}_{\mathrm{(\lc,\pi^{-, +})\leftarrow \Sigma_\mathrm{c}^{0,++}}}(\Dphi)$). 
These generated templates were then normalised by a factor which accounted for the expected contribution of $\lc\leftarrow \Sigma_\mathrm{c}^{0,++}$ in the analysed data sample. Such a factor was evaluated from ALICE measurements of feed-down contribution to $\lc$ production from $\Sigma_\mathrm{c}^{0,++}$ ~\cite{ALICE:2021rzj}.

The fully-corrected, per-trigger azimuthal correlation distribution of prompt $\lc$ baryons and charged particles was obtained as summarised in Eq.~\ref{eq:FinalCF}:
\begin{equation}
    \label{eq:FinalCF}
    \begin{split}
&\frac{1}{N_{\lc}}\frac{\mathrm d N^\mathrm{assoc}(\Dphi)}{\rm d\Dphi} = \\
&\frac{1}{f_{\mathrm{prompt}}}\left[b_{\mathrm{bias}}(\Dphi)\,\Tilde{C}_{\mathrm{inclusive}}(\Dphi) - (1-f_{\mathrm{prompt}})\,\Tilde{C}^\mathrm{MC}_\mathrm{feed-down}(\Dphi) \right] 
- \frac{\sigma(\lc\leftarrow \Sigma_\mathrm{c}^{0,++})}{\sigma(\lc)}\,\Tilde{C}^\mathrm{MC}_{\mathrm{(\lc,\pi^{-, +})\leftarrow \Sigma_c^{0,++}}}(\Dphi). 
\end{split}
\end{equation}

\subsection{Extraction of correlation properties}

In order to quantify its properties, the $\lc$ azimuthal-correlation distribution was fitted using the following function, already exploited and motivated in previous D-meson azimuthal correlation studies~\cite{ALICE:2019oyn}: 
\begin{equation}
\label{equ:Fit}
    f(\Dphi) = b + \frac{Y_{\rm NS}\times\beta}{2\alpha \Gamma(1/\beta)}\times e^{-\left(\frac{\Dphi}{\alpha}\right)^\beta} + \frac{Y_{\rm AS}}{\sqrt{2\pi}\sigma_{\rm AS}}\times e^{-\frac{(\Dphi - \pi)^2}{2\sigma^2_{\rm AS}}}.
\end{equation}
The function is composed of a generalised Gaussian component for the description of the near-side (``NS'') peak, a Gaussian component for the away-side (``AS'') peak, and a constant term (baseline, $b$) to account for the flat contribution that lies beneath the two correlation peaks. Specifically, the yields of the peaks on the near and away sides ($Y_{\rm NS}$ and $Y_{\rm AS}$, respectively) were calculated by integrating the components representing each correlation peak. The near- and away-side peak widths were characterised by the parameters $\alpha\sqrt{\Gamma(3/\beta)/\Gamma(1/\beta)}$ (the square root of the variance of the generalised Gaussian) and $\sigma_{\rm AS}$, respectively. To grant sufficient stability to the fit, the $\beta$ parameter of the generalised Gaussian was set to the value obtained by fitting the correlation distribution predicted by PYTHIA~8 simulations with the CR-BLC Mode 2 tune~\cite{Christiansen:2015yqa}. The typical $\beta$ parameter values varied within the range $0.7-1.9$, decreasing for increasing $\lc$ transverse momentum.
The mean values of the near-side generalised Gaussian and of the away-side Gaussian were also fixed at  $\Delta\varphi = 0$ and $\Delta\varphi = \pi$, respectively. All the other fit parameters were kept free. The consistency of the model parameterisation for the $\beta$ values was evaluated by employing the Von Mises function~\cite{upton2008dictionary} as an alternative to fit the azimuthal correlation distribution. The two fitting procedures gave compatible results, and the selection of the fit function in Eq.~\ref{equ:Fit} was motivated by the need for a consistent comparison with the results from prompt D-meson azimuthal correlation studies~\cite{ALICE:2021kpy}, as detailed in Section~\ref{sec:Results}.

%% file: Systematics.tex
\section{Systematic uncertainties}
\label{sec:Systematics}

The measured azimuthal correlation distribution is affected by several systematic uncertainties related to specific steps of the analysis procedure, or to assumptions introduced for performing the measurement. In this section, the method employed to estimate each source of systematic uncertainty, using the same strategy discussed in Ref.~\cite{ALICE:2021kpy}, is briefly outlined.

A systematic uncertainty arises from the evaluation of $S_{\mathrm{peak}}$ and $B_{\mathrm{peak}}$ from the fit to the $\Lc$-baryon invariant-mass distributions. It was assessed by modifying the fitting procedure, including variations of the background fit function, histogram binning, fit range, and fixing the Gaussian parameters to the values obtained from Monte Carlo simulations. The resulting uncertainty ranged from 1 to 2\% depending on the $\pt$ interval, with no dependence on $\Delta\varphi$.

A systematic uncertainty ranging from 0.5 to 2\%, depending on the trigger and the associated particle $\pt$, was introduced due to a potential dependence of the shape of the background correlation distribution on the invariant-mass value of the trigger $\Lc$ baryon. This uncertainty was assessed by evaluating the sideband correlation distribution, $\Tilde{C}_{\rm{sidebands}}(\Dphi, \Delta\eta)$, in different invariant-mass ranges with respect to the default 4-8$\sigma$ regions. No azimuthal-angle dependence was observed for this uncertainty.

The evaluation of the associated-particle reconstruction efficiency from Monte Carlo simulations introduced an additional systematic uncertainty, estimated by applying different quality selection criteria on the sample of reconstructed tracks, such as removing or tightening the request on the minimum number of ITS clusters, requiring a hit on at least one of the two SPD layers, or varying the request on the minimum number of space points reconstructed in the TPC. The uncertainty related to the ITS–TPC track matching efficiency was considered as well. An uncertainty of 4\% was estimated, with no significant trend in $\Delta\varphi$.

A systematic uncertainty affecting the $\Lc$ reconstruction efficiency, attributed to possible discrepancies in the distributions of the topological variables in Monte Carlo and data, was estimated by repeating the analysis exploiting a set of tighter and looser scores for the BDT model used for the selection of $\Lc$ candidates. An uncertainty ranging from 0.5 to 2\% was assigned, depending on the $\pt$ interval, with no $\Delta\varphi$ dependence.

An additional systematic uncertainty is related to the evaluation of the residual contamination from secondary particles via Monte Carlo studies. To determine its value, the analysis was repeated by testing different DCA selections in the $xy$ plane, ranging from 0.1 to 2.4 cm, and re-evaluating the purity correction of primary tracks for each variation. This resulted in a maximum, $\Delta\varphi$-independent, systematic uncertainty of 1 to 2\% on the azimuthal-correlation distribution.

In addition to the above-mentioned uncertainties, which all affect the azimuthal correlation distribution as scale factors, some $\Delta\varphi$-dependent systematic uncertainties are present, related to the feed-down subtraction procedure and to the removal of the contamination of soft pions from $\Sigma_{\rm c}^{0,++}$ decays. 

To assess the uncertainty related to the subtraction of the beauty feed-down contribution to the measured correlation distribution, different tunes of the PYTHIA~8 event generator were used to estimate the correlation distribution of feed-down $\Lc$ triggers. In addition, $f_{\rm prompt}$ was varied following the procedure described in Ref.~\cite{ALICE:2019nxm}. A $\Delta\varphi$-dependent and asymmetric uncertainty was obtained, reaching a maximum of 5\% in the near-side peak region and decreasing to 0 in the region outside the correlation peaks. 
Moreover, a $\Delta\varphi$-dependent symmetric uncertainty on the $b_{\rm{bias}}(\Delta\varphi)$ was considered, to take into account a possible over- or underestimation of that correction factor, as detailed in~\cite{ALICE:2019oyn}. This uncertainty reached a maximum of 2\% for $\Delta\varphi \approx 0$.

The subtraction procedure of soft pions produced in $\Sigma_{\rm c}^{0,++} \rightarrow \Lc\pi^{-,+}$ decays relies on the measured fraction of $\Lc$ produced from $\Sigma_{\rm c}^{0,++}$ decays, which is affected by statistical and systematic uncertainties~\cite{ALICE:2021rzj}. The subtraction procedure was repeated by shifting by $\pm 1\sigma_{\rm total}$ the value of such a fraction, and the resulting systematic uncertainty was quantified based on the impact on the corrected azimuthal correlation distributions. A maximum value of 0.5\% for $\Delta\varphi \approx 0$ was obtained.

The systematic uncertainty values from the mentioned sources impacting the azimuthal correlation distribution are summarised in Table~\ref{systematic_table}. The total systematic uncertainty for each $\Delta\varphi$ bin in the correlation distribution was derived by summing the contributions mentioned above in quadrature.

\begin{table}[th]
  \caption{Systematic uncertainty contributions influencing the azimuthal correlation distribution along with their typical values. If not specified, the uncertainties do not depend on $\Delta\varphi$.}
\centering
 \begin{tabular}{|l|c|}
 \hline
 \textbf{Source} & \textbf{Uncertainty} \\
 \hline
 Yield extraction & 1--2\%  \\
 Background $\Delta\varphi$ distribution & 0.5--2\%  \\
 Associated-track reconstruction efficiency & 4\%  \\
 $\Lc$-baryon reconstruction efficiency & 0.5--2\%  \\
 Primary-particle purity & 1--2\%  \\
 Feed-down subtraction & $\leq$5\%, $\Delta\varphi$-dependent \\
 Selection bias for feed-down contribution & $\leq$2\%, $\Delta\varphi$-dependent \\
 Soft-pion subtraction & $\leq$0.5\%, $\Delta\varphi$-dependent \\
 \hline
 \end{tabular}

 \label{systematic_table}
\end{table}

The systematic uncertainties impacting the near- and away-side peak observables, i.e. peak yields and widths, were assessed based on the following contributions: 
(i) the total $\Delta\varphi$-independent systematic uncertainty, which affects the correlation distribution as a scale factor, and hence impacts the near- and away-side peak yield values by the same relative amount; (ii) the $\Delta\varphi$-dependent uncertainty on the correlation distribution, whose impact on the peak observables was addressed by repeating the fits after shifting all the points of the correlation distribution upwards and downwards by $1\sigma$; (iii) the uncertainty related to the fit configuration, in particular on the choice of the baseline value and of the generalised Gaussian $\beta$ parameter. The impact of this uncertainty was evaluated by repeating the fit after determining the baseline position using alternative $\Dphi$ ranges, as well as leaving the $\beta$ parameter free, and calculating the root-mean-square of the peak observables evaluated from the variations.
Among the above uncertainty sources, the choice of the baseline value represents the dominant contribution to the peak-observable uncertainties. 
The overall systematic uncertainty on the peak yields was determined by summing in quadrature the contributions from (i), (ii), and (iii). For the widths, being insensitive to scale factors, only the sum in quadrature of the contributions from (ii) and (iii) was considered. The ranges of the overall systematic uncertainties on these observables are 8--29\% for the near-side yields, 6--30\% for the near-side widths, and 12--54\% for the away-side yields, depending on the transverse momentum range.
In the calculation of the ratios of the $\Lc$ baryons and D mesons associated yields and widths, the mentioned uncertainties were propagated as uncorrelated with the exception of those related to the tracking efficiency of the associated particles and to the beauty feed-down subtraction. The former was assumed to be fully correlated and cancels out in the ratios; for the latter, for each observable, the ratio was recalculated varying coherently the assumptions involved in the feed-down subtraction for both  $\Lc$ baryons and D mesons azimuthal correlation functions and the uncertainty was estimated from the envelope of the values obtained.

%% file: Results.tex
\section{Results}
\label{sec:Results}
The azimuthal correlation distributions of $\Lc$ baryons with charged particles in pp collisions at $\sqrt{s}=13$ TeV were measured for the three $\Lc$-baryon $\pt$ intervals $3 < \ptL < 5$~\GeVc, $5 < \ptL < 8$~\GeVc, and $8 <\ptL< 16$~\GeVc as well as for the associated-particle $\pt$ ranges $\ptass > 0.3$~\GeVc, and its sub-intervals $0.3 < \ptass < 1$~\GeVc and  $\ptass > 1$~\GeVc.

\subsection{Analysis results and comparison with D-meson and charged particle correlations}
\begin{figure} 
    \centering
    \includegraphics[scale=0.8]{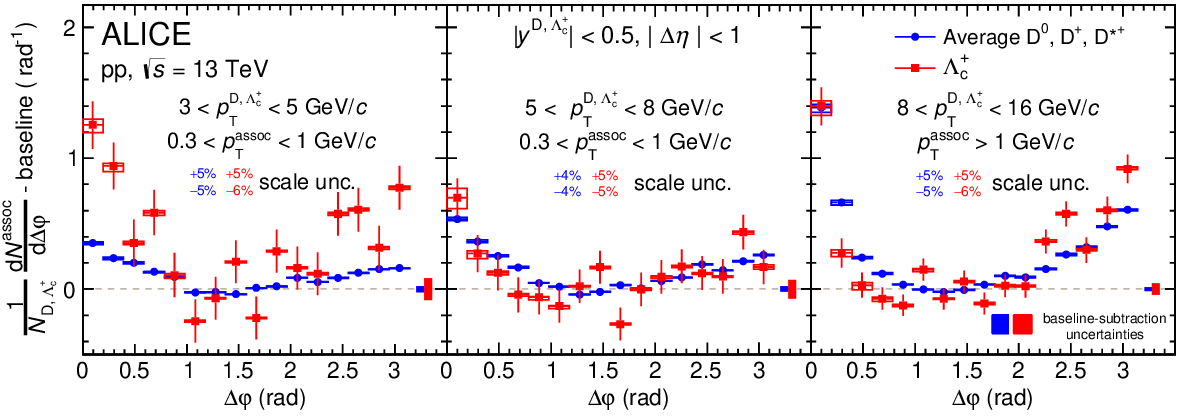}
    \caption{Examples of azimuthal-correlation distributions of $\Lc$-baryons with associated particles~(red markers) after the baseline subtraction in pp collisions at $\sqrt{s}$ = 13 TeV, compared to the average of the azimuthal-correlation distributions of $\Dzero$, $\Dplus$, and $\Dstar$ mesons with associated particles measured by ALICE~(blue markers)~\cite{ALICE:2021kpy}, for the $\pt$ intervals 3 $< \ptDL <$ 5~\GeVc and 0.3 $<\ptass<$ 1~\GeVc, 5~$< \ptDL <$~8~\GeVc and 0.3 $<\ptass<$ 1~\GeVc, and 8 $<\ptDL<$ 16~\GeVc and $\ptass>$ 1~\GeVc(from left to right). Statistical and $\Delta \varphi$-dependent systematic uncertainties are shown as vertical error bars and boxes, respectively, and $\Delta \varphi$-independent uncertainties are written as text. The uncertainties from the subtraction of the baseline are displayed as boxes at $\Delta \varphi>\pi$.}
    \label{fig:results:dPhiComparison_DvsLc}
\end{figure}

The correlation distributions of $\Lc$ baryons and charged particles are compared to the results obtained considering the average of $\Dzero$, $\Dplus$, and $\Dstar$ mesons as triggers~\cite{ALICE:2021kpy} after the subtraction of the baseline, whose values were found to be consistent between the two trigger-particle cases for all the studied $\pt$ intervals. This allows for quantifying possible differences in the correlation peak features related to the different hadronisation of the charm quark into baryons and mesons.
The comparison of the baseline-subtracted correlation distributions for a selection of the studied transverse momentum intervals is presented in Fig.~\ref{fig:results:dPhiComparison_DvsLc}.
Because of the smaller $\lc$ signal, the reduced reconstruction efficiency and the larger combinatorial background under the invariant mass peaks with respect to the D-meson analyses, the baryon measurements are affected by larger point-by-point statistical fluctuations. 
The shape of the angular correlation distribution between $\lc$ and charged particles points toward a difference compared to the correlation between D mesons and charged particles in some $\pt$ regions probed by the measurement. More specifically, indications of a larger amount of correlation pairs emerge in both near- and away-side correlation peaks in the transverse momentum interval $3<\ptL<5$~\GeVc, and for associated particles in $0.3<\ptass<1$~\GeVc. For more energetic charm hadrons, starting from trigger-particle $\pt>5$~\GeVc, the agreement between the angular distributions of the two charm hadrons is significantly improved and the correlation functions are found to be compatible. 
At high associated-particle $\pt$, the azimuthal correlation distribution of the two charm hadrons exhibits a similar shape, despite hints of a tension being present in the peak regions, in particular for the near-side.

\begin{figure}
    \centering
    \includegraphics[scale=0.78]{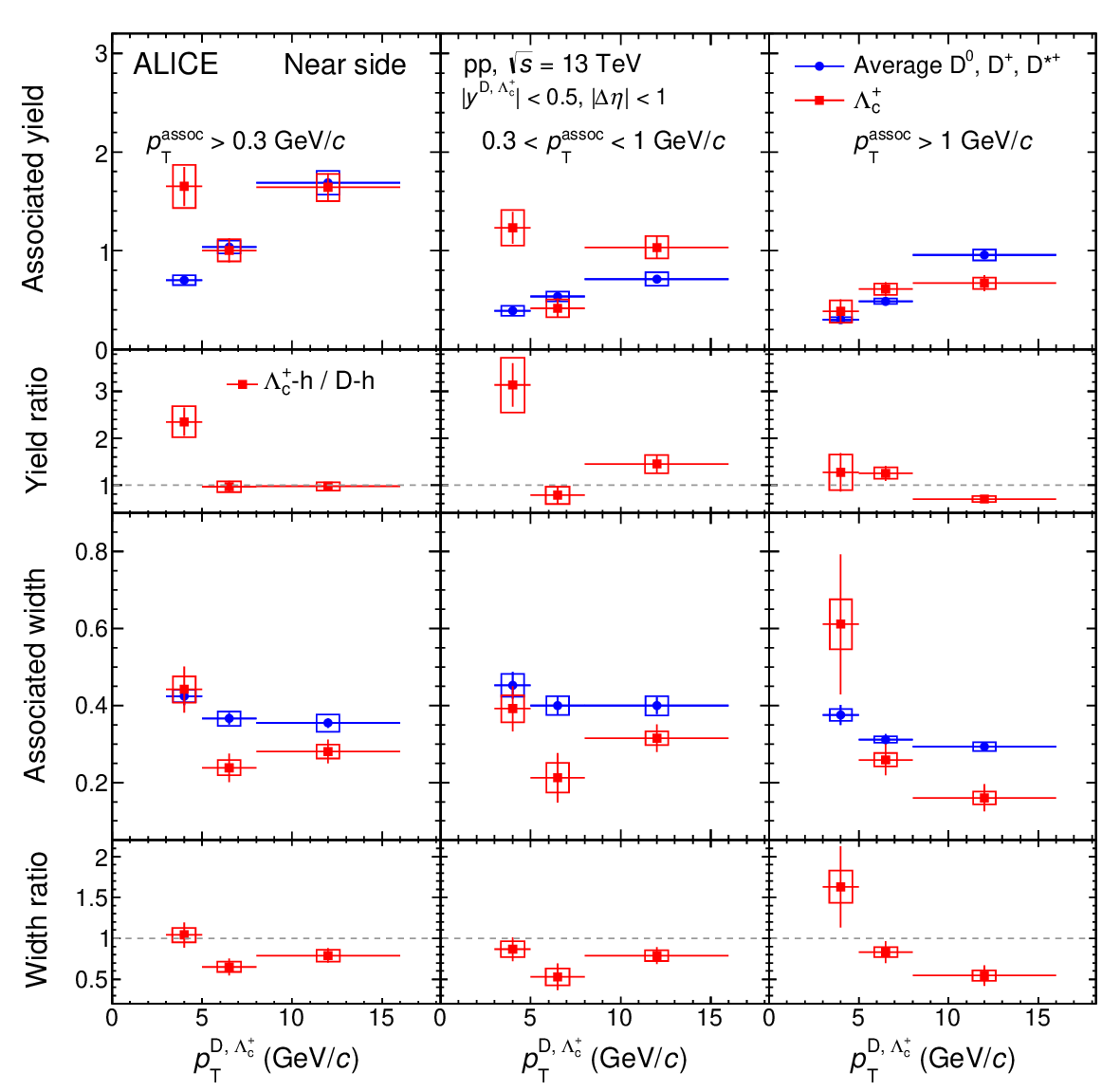}
    \caption{Near-side peak yields (first row) and widths (third row) obtained from the fit to the azimuthal correlation distributions of $\Lc$ with charged particles after the baseline subtraction in pp collisions at $\sqrt{s}=13$ TeV. The measurements are compared to D-meson average results by ALICE in the same collision system~\cite{ALICE:2021kpy}. The ratios of $\Lc$-baryon to D-meson near-side peak observables are shown in the second and fourth rows for yields and widths, respectively.}
    \label{fig:results:NSvsD}
\end{figure}

From the fit to the $\Lc$ azimuthal correlation distributions, more quantitative insights can be gained into the evolution of the correlation peak features as a function of the transverse momentum of the trigger and the associated particles. In Fig.~\ref{fig:results:NSvsD}, the near-side yields and widths are compared to the corresponding values for D-meson correlation distributions~\cite{ALICE:2021kpy}. 
In the near-side region, this comparison suggests a larger associated peak yield for $\Lc$-triggered correlations compared to D-meson results in the charm-hadron transverse momentum interval $3<\ptDL<5$~\GeVc. This effect appears to be more pronounced for associated particles with transverse momentum $0.3<\ptass<1$~\GeVc, where a deviation of 2.7$\sigma$ is observed, and also appears when considering the $\ptass>0.3$~\GeVc range to which the associated particles in the $0.3<\ptass<1$~\GeVc range give a large contribution for this $\ptL$ interval. This difference is not present for higher $\ptass$ values.
For increasing charm-hadron transverse momentum, i.e. $\ptDL>5$~\GeVc, the near-side yields values from the azimuthal correlation distributions of $\lc$ baryons with charged particles are compatible with those of D-mesons in the $\ptass>0.3$~\GeVc range.
Considering the two associated-particle $\pt$ ranges $0.3 < \ptass < 1$~\GeVc and $\ptass > 1$~\GeVc, some point-by-point differences are present in the yield values, but they are not statistically significant and do not show systematic trends with the current uncertainties.
For $3<\ptDL<5$~\GeVc the values of the near-side widths for the $\Lc$- and D meson-triggered correlations are observed to be similar for all the three $\ptass$ intervals, while for $\ptDL>5$~\GeVc a statistically significant sharpening of the near-side peak beyond the 99\% confidence level for $\ptass > 0.3$ GeV/$c$, is observed. 

The hint of larger associated peak yields at low $\pt$ for the $\Lc$-triggered correlations compared to D-meson-triggered correlations, and similar baseline values among the two cases, is particularly relevant in connection with the ALICE measurement of $\Lc/\Dzero$ production yield ratios as a function of the event multiplicity in pp collisions at \s = 13 TeV~\cite{ALICE:2021npz}. In that case, larger values of the baryon-to-meson ratio were observed for $2 < \pt < 12$ GeV/$c$ in collisions with higher charged-particle multiplicity. In a scenario where charm-jet fragmentation and soft-particle production are considered uncorrelated, the above observation from the comparison of the baseline and near-side yield values may suggest that the measured increase of $\Lc/\Dzero$ yield ratio with multiplicity could be related to an increased production of associated particles in the jets containing a $\Lc$, rather than to a higher probability of forming a $\Lc$ in events with a larger multiplicity of uncorrelated particles. 

\begin{figure}
    \centering
    \includegraphics[scale=0.78]{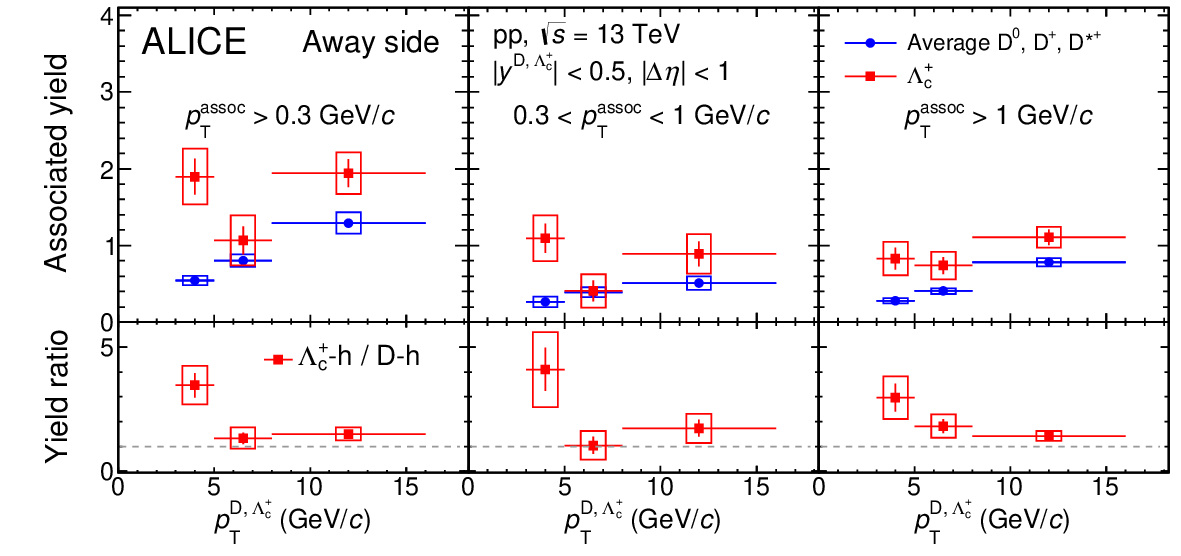}
    \caption{Away-side peak yields (first row) obtained from the fit to the azimuthal correlation distributions of $\Lc$ and charged particles after the baseline subtraction in pp collisions at $\sqrt{s}=13$ TeV. The measurements are compared to D-meson average results by ALICE in the same collision system~\cite{ALICE:2021kpy}. The ratios of $\Lc$-baryon to D-meson away-side peak yields are shown in the second row.}
    \label{fig:results:ASvsD}
\end{figure}

Albeit having a looser connection with charm fragmentation and hadronisation than the near-side, the away-side region could retain information on the fragmentation of the recoil parton and provide additional constraints for the interpretation of the charm shower in the near-side.
The away-side peak yields, shown in Fig.~\ref{fig:results:ASvsD}, are characterised by larger statistical and systematic uncertainties compared to the near-side yields. In correlations of $\Lc$ baryons with charged particles, larger values of the away-side yields with respect to those of D mesons are generally obtained for the kinematic intervals studied. Similarly to the near-side case, the largest difference is observed for $3<\ptL<5$ GeV/$c$.
An increased production of low-transverse momentum associated particles, both collimated with the trigger $\Lc$ baryon and in the opposite azimuthal direction, could be explained by a possible softer fragmentation of the charm quark when hadronising into a $\Lc$ baryon rather than a D meson. This is supported by measurements of the longitudinal momentum fraction of $\Lc$ baryons reconstructed in jets~\cite{ALICE:2023jgm}. More specifically, a softer fragmentation would imply that, on average, the original charm quark retains a smaller fraction of its initial energy, leaving more phase space for the production of additional charged particles in the near-side region. Furthermore, a larger initial energy of the charm quark typically corresponds to a larger $Q^2$ of the hard-parton scattering process. This, in turn, would on average lead to an increased energy of the other (anti)charm quark produced in the scattering, potentially enhancing particle production on the away-side as well, as also observed from PYTHIA~8 predictions.
No comparison of the away-side peak widths between $\Lc$-triggered and D-meson-triggered correlation measurements is reported, as the large uncertainties affecting the measurement of this observable prevent drawing significant conclusions.

\subsection{Comparison with model predictions}
The peak properties of the $\Lc$-- charged particle azimuthal correlation distributions were also compared to the state-of-the-art Monte Carlo simulations and model predictions, which include different treatments for the charm-quark parton showering process and hadronisation mechanisms. Among the available predictions, of particular interest were those succeeding in the description of the charm $\Lc/\Dzero$ ratios measured by ALICE~\cite{ALICE:2022exq,ALICE:2021rzj}. In particular, predictions from the PYTHIA~8 simulations including colour reconnection mechanisms beyond the leading-colour approximation (CR-BLC)~\cite{Christiansen:2015yqa} are considered, which include new colour-junction topologies which lead to baryon formation more frequently, thus increasing the production of $\Lc$ baryons.
The three existing CR-BLC configurations of PYTHIA~8 provide very similar predictions for the azimuthal-correlation distributions of $\Lc$ baryons and charged particles.
For this reason, only the predictions from the mode 2, advised by the authors as the default CR-BLC tune, are included in the comparison with the measurements. This choice also ensures consistency with the data-to-model comparison reported in Ref.~\cite{ALICE:2023jgm} for the measurement of charm-tagged jets.

The PYTHIA~8 predictions with the Monash tune~\cite{Sjostrand:2014zea}, where the quark fragmentation is tuned on e$^+$e$^-$ and ep measurements, and the POWHEG+PYTHIA~8~\cite{Nason:2004rx,Frixione:2007vw} predictions, based on NLO hard scattering matrix elements, are also included in the comparison. The expectations from the JETSCAPE model, with Hybrid hadronisation~\cite{JETSCAPE:2019udz, Han:2016uhh}, are also considered. They include the possibility of baryon formation via quark coalescence in addition to the vacuum fragmentation as implemented in the PYTHIA~8 framework.
Using the same kinematic prescriptions and considering the same associated particle species as for the data analysis, the model predictions for the correlation peak observables were extrapolated by fitting the simulated correlation distribution with Eq.~\ref{equ:Fit}, where the $\beta$ parameter was left unconstrained, and the baseline was computed as the minimum of the correlation distributions. In order to include the uncertainty on the peak observables originated by the baseline definition, a systematic uncertainty was computed by fixing the baseline as the weighted average of the two lowest points of the azimuthal correlation function and propagated to the final observables.

\begin{figure}
    \centering
    \includegraphics[scale=0.78]{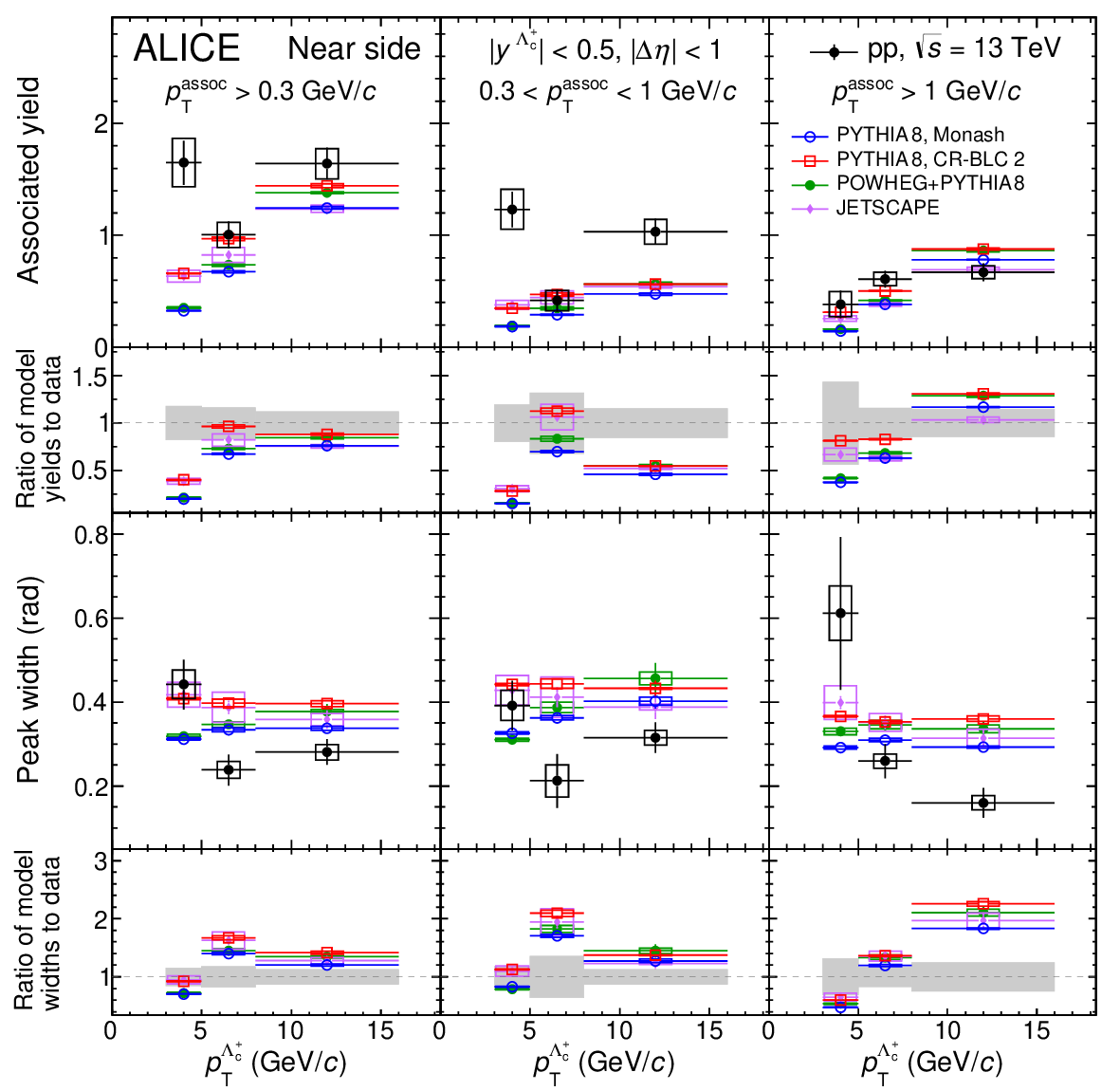}
    \caption{Near-side peak yields (first row) and widths (third row) from the fit to the $\Lc$-charged particle azimuthal correlation distributions after the baseline subtraction, compared to simulations from PYTHIA~8 with Monash tune~\cite{Sjostrand:2014zea}, POWHEG+PYTHIA~8~\cite{Nason:2004rx,Frixione:2007vw}, PYTHIA~8 with CR-BLC mode 2~\cite{Christiansen:2015yqa}, and JETSCAPE with hybrid hadronisation~\cite{JETSCAPE:2019udz, Han:2016uhh}. The ratios of model predictions to data measurements for the yield (width) values are shown in the second (fourth) row. In these rows, model statistical and systematic uncertainties are shown as vertical error bars and boxes, respectively, while data total uncertainties are displayed as a solid grey band.}
    \label{fig:results:ppLcMC_NS}
\end{figure}

The near-side peak observable evolution with increasing transverse momentum of the $\lc$ particle for the various models is illustrated in Fig.~\ref{fig:results:ppLcMC_NS} and compared to data results, for the $\ptass > 0.3$~\GeVc interval and in the two sub-intervals studied. 
The first row shows the measured yield values, while the second row shows the ratios of model predictions for the yields with respect to data. In the ratios, the statistical and systematic uncertainties affecting the models are shown as error bars and boxes, respectively, while the data statistical and systematic uncertainties are summed in quadrature, and represented as a solid grey band.
The comparison points toward an underestimation of the measured near-side yields by all the model predictions for low-momentum $\lc$-baryons and associated particles in $0.3<~\ptass~<1.0$~\GeVc. A generally better description is provided by the models starting from $\ptL>5$~\GeVc for $\ptass>0.3$~\GeVc, although some tensions are observed for the breakdown into the two associated-particle transverse momentum intervals $0.3<~\ptass~<1.0$~\GeVc and $\ptass~>1.0$~\GeVc, with the models providing 50\% lower yields on average at high-$\ptL$ and small $\ptass$. All the model predictions are characterised by an increasing trend of the near-side yield with the trigger-particle transverse-momentum. For all the kinematic regions, PYTHIA~8 Monash tune and POWHEG+PYTHIA~8 yield expectations are lower compared to other models, with POWHEG+PYTHIA~8 providing slightly larger yields with increasing $\Lc$-baryon energy, as already observed in D-meson measurements. 
The largest near-side yield values are given by the PYTHIA~8 with CR-BLC mode 2 tune. The JETSCAPE model with hybrid hadronisation shows a different trend: while at low-$\ptL$ compatible values are obtained with the expectations of the PYTHIA~8 with CR-BLC tune, with increasing $\pt$ of the trigger particle a milder increase of the near-side yields is observed for JETSCAPE. This leads to a smaller associated yield of charm-jet and an increased difference between the two models. 
The results from JETSCAPE allow us to obtain some indications of the possible contribution of the coalescence hadronisation mechanism to the tension between $\Lc$- and D-meson-triggered correlations. In general, the momentum of a charm baryon produced by coalescence is expected to be larger than that of a baryon from the fragmentation of a charm quark with the same energy. Therefore, for a given $\lc$ transverse momentum, in the coalescence scenario, the near-side peak yield would correspond to that produced by a lower-energy charm quark, and would thus be reduced. In addition, this mechanism foresees the presence of light-flavour quarks to combine with the charm quark to form the final charm-baryon state, therefore preventing those quarks from forming additional associated particles. If the coalescence occurs with light quarks that would otherwise produce particles entering the charm-jet cone, this might also result in a reduction of the near-side peak yield. On the other hand, the charm jet might be affected by the coalescence of the partons produced in the charm showering with nearby partons emerging from the underlying event, making it broader and more energetic.
Predictions by JETSCAPE with hybrid hadronisation seem to disfavour sizeable effects of coalescence on the correlation distributions. Further comparisons with other models that include hadronisation via coalescence and that can also reproduce the baryon-to-meson production cross section ratio~\cite{ALICE:2020wfu, ALICE:2022exq}, such as Catania~\cite{Minissale:2020bif} and QCM~\cite{Song:2018tpv}, could provide additional insights on this topic.

The third row of Fig.~\ref{fig:results:ppLcMC_NS} shows the measured near-side width values, while the fourth row shows the ratios of model predictions for the widths with respect to data. This observable is not described properly by the models: while for $3<\ptL<5$ \GeVc the agreement with the measurement is satisfactory, for larger values of $\ptL$ all the models tend to overestimate the width values. Two separate trends can be highlighted among the model predictions:  while PYTHIA~8 with CR-BLC mode 2 tune and JETSCAPE foresee a flatter dependence of widths as a function of the trigger $\pt$, with hints of a decreasing trend for increasing $\ptL$, PYTHIA~8 with Monash tune and POWHEG+PYTHIA~8 hint towards larger widths with increasing momentum of the $\Lc$.

\begin{figure}
    \centering
    \includegraphics[scale=0.77]{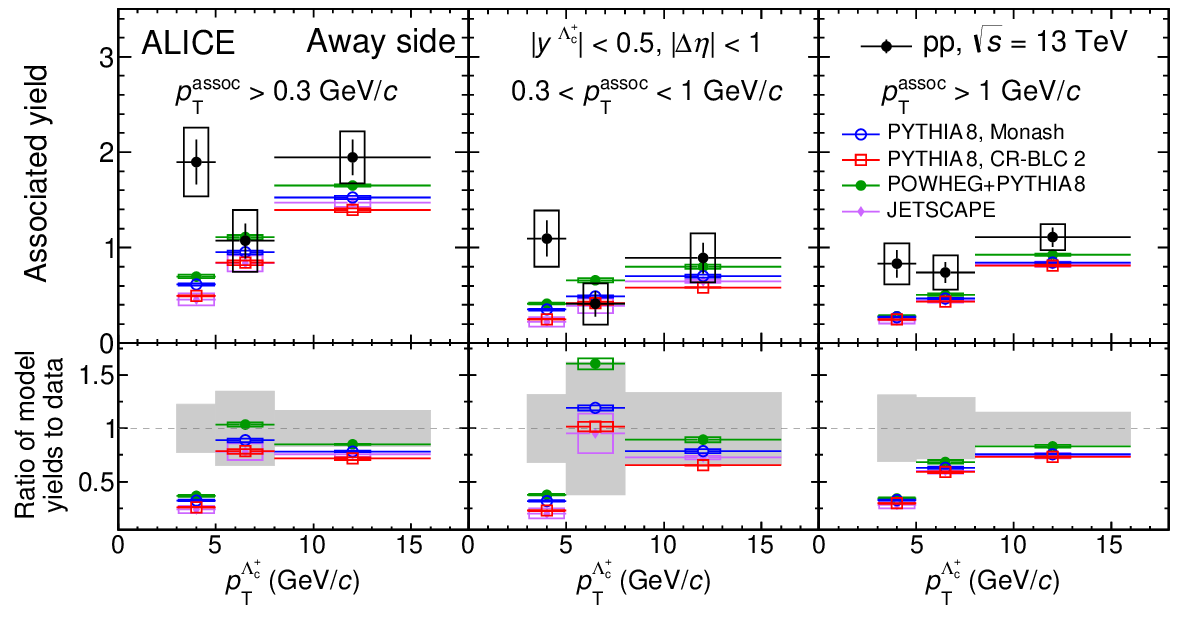}
    \caption{Away-side peak yields (first row) from the fit to the $\Lc$-charged particle azimuthal-correlation distributions after the baseline subtraction, compared to simulations from PYTHIA~8 with Monash tune~\cite{Sjostrand:2014zea}, POWHEG+PYTHIA~8~\cite{Nason:2004rx,Frixione:2007vw}, PYTHIA~8 with CR-BLC mode 2~\cite{Christiansen:2015yqa}, and JETSCAPE with hybrid hadronisation~\cite{JETSCAPE:2019udz, Han:2016uhh}. The ratios of model predictions to data measurements for yield values are shown in the second row. In this row, model statistical and systematic uncertainties are shown as vertical error bars and boxes, respectively, while data total uncertainties are displayed as a solid grey band.}
    \label{fig:results:ppLcMC_AS}
\end{figure}

Concerning the model description of the away-side peak yields, shown in Fig.~\ref{fig:results:ppLcMC_AS}, similar conclusions can be drawn to those for the near-side peak.
The trigger and associated particle low-momentum region appears to be underestimated, though data uncertainties are larger, while for larger $\ptL$ a qualitatively good agreement within uncertainties is observed between data and predictions. It is interesting to observe that a reversed hierarchy characterises the away-side description.  The PYTHIA 8 with CR-BLC mode 2 tune predicts the smallest yields, compatible also with the JETSCAPE model, closely followed by PYTHIA~8 with Monash tune and POWHEG+PYTHIA~8.

It is worth noting that while PYTHIA~8 (for all the tunes considered) and POWHEG+PYTHIA~8 predictions describe the peak properties of the azimuthal correlations of D mesons with charged particles at midrapidity~\cite{ALICE:2021kpy} within uncertainties, the same models are not able to provide an accurate description of the $\Lc$-charged particle correlation measurements, at least for the low-momentum range.

\subsection{PYTHIA 8 predictions with additional charm-baryon resonant states}
\begin{figure}
    \centering
    \includegraphics[scale=0.45]{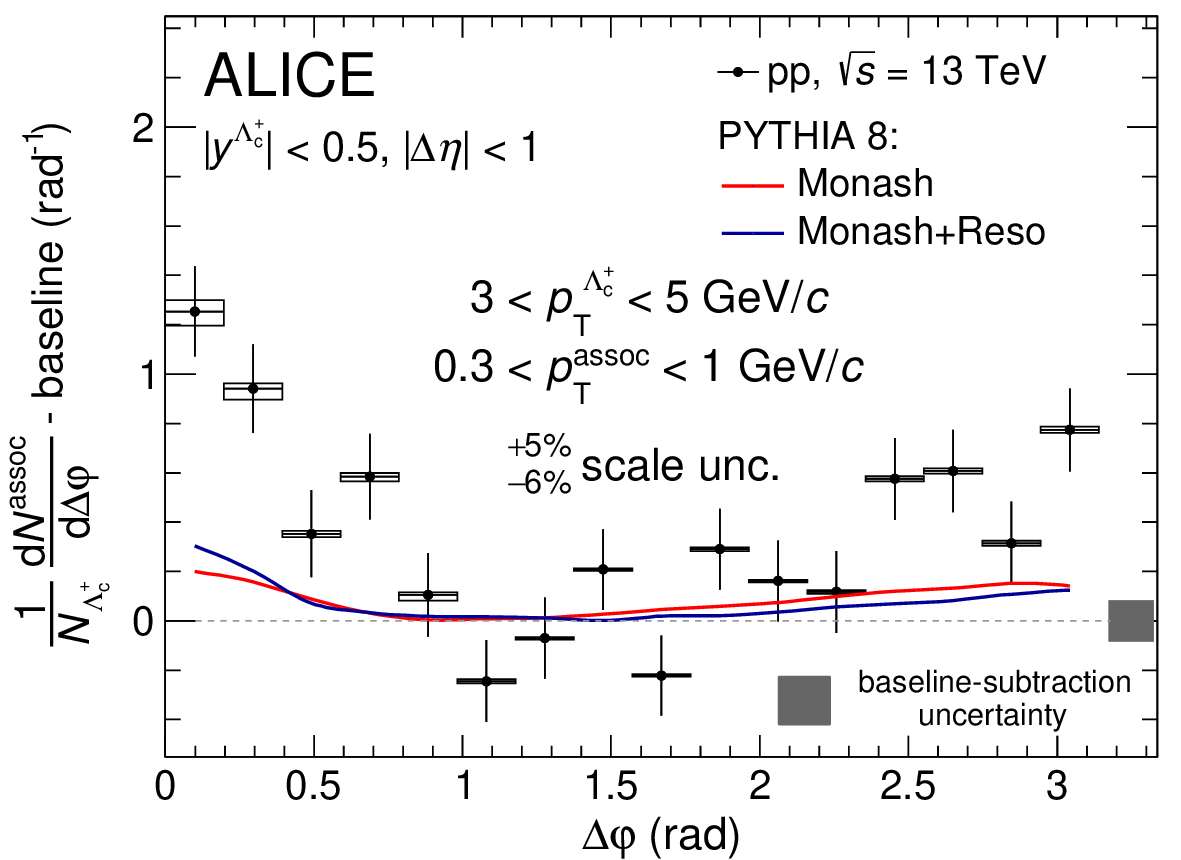}
    \includegraphics[scale=0.335]{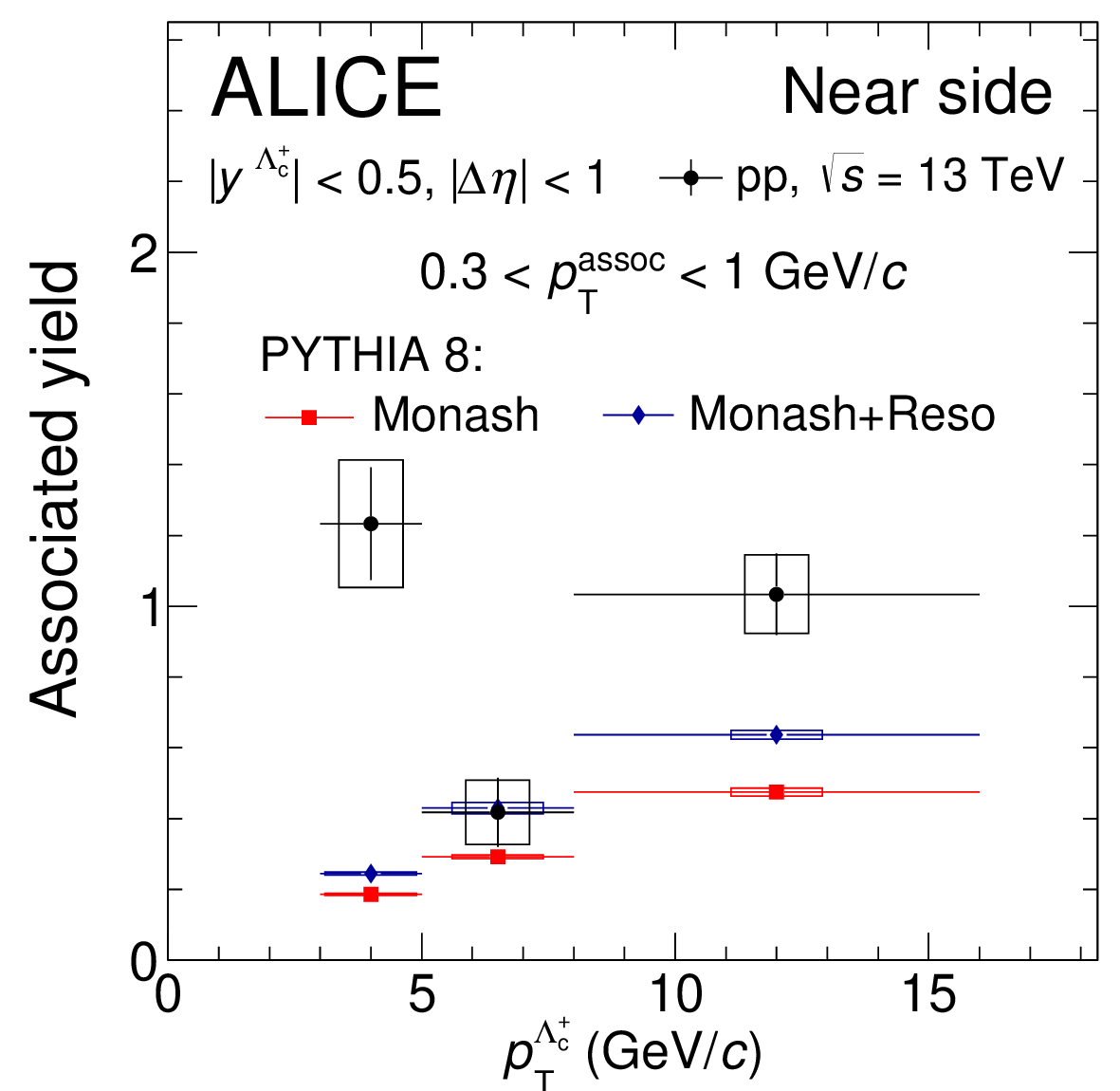}
    \caption{Left: comparison between the measured $\Lc$-charged particle azimuthal correlation function and the Monte Carlo correlation templates obtained from PYTHIA~8, Monash tune and by including SHM+RQM charm baryon states (see text for details) for $3 < \ptL < 5\,\gev/c$ and for 0.3 $<\ptass<$ 1 GeV/$c$. Right: the near-side yields from data compared to the same model predictions for 0.3 $<\ptass<$ 1 GeV/$c$.}%
    \label{fig:shmrqm}
\end{figure}

The presence of an additional contribution to $\Lc$ production yields from the decay of heavier and yet-unobserved charm-baryon resonant states, as predicted by the SHM+RQM model~\cite{He:2019tik}, is one of the proposed explanations for the enhancement in $\lc/\Dzero$ production yield ratio observed at low and intermediate $\pt$ in pp collisions with respect to ${\rm e}^+{\rm e}^-$ ~\cite{ALICE:2023sgl}.
In order to probe modifications to the $\Lc$-charged particle correlation shape induced by this possible additional charm feed-down contribution, a dedicated model based on PYTHIA 8 simulations was developed, by taking into account the decay kinematics of a given set of charm-baryon resonant states decaying into $\lc$ and charged particles, and by evaluating the angular separation between the $\Lc$ and the other decay products. As it was done in the SHM+RQM model for
estimating the $\lc$ production cross section~\cite{He:2019tik}, the decay simulation was performed considering ``average'' baryon states, each of them aggregating all the states with different spin values of the same particle species~($\Lc$, $\mathrm{\Sigma_\mathrm{c}}$, $\mathrm{\Xi_\mathrm{c}}$, and $\mathrm{\Omega_\mathrm{c}}$).
Additionally, because no precise prescriptions are available about the decay chain of such states, the decay channels and the branching ratios of the average baryon states were modelled on the observed decays of the corresponding ground-state charm-baryon measured decay channels. 
The decays of the resonant states were handled by the PYTHIA 8 Monte Carlo generator.
For each transverse momentum interval and for each additional charm-baryon resonant state defined, the resulting azimuthal correlations between the feed-down $\lc$ and the other charged particles produced by the decay of the charm-baryon resonance were evaluated. To properly account for the SHM expected resonance $\pt$ spectrum, the per-trigger $\lc$-charged particle correlation distributions from the additional SHM+RQM states were corrected by a factor including a modified fragmentation function of charm quarks in these states depending on their mass, as described in~\cite{He:2019tik}. The azimuthal correlation templates from the decay of each of the additional states were also reweighted to the expected fraction of feed-down $\lc$ yield from that state, with respect to the total amount of produced $\lc$. Similarly, the azimuthal correlation templates of prompt $\lc$ with charged particles were also scaled to the yields expected from SHM+RQM. Finally, the per-trigger reweighted $\lc$-charged particle distributions from the decay of the SHM+RQM average states were summed to the per-trigger azimuthal correlation distributions between $\Lc$ and other charged particles generated with the PYTHIA~8, Monash tune.
This model, called PYTHIA~8 Monash+Reso, simulates the feed-down contribution from the augmented charm-baryon states only for the trigger $\Lc$ baryons, while the fragmentation and hadronisation of the other charm quark are left unchanged with respect to PYTHIA~8 with the Monash tune.
An example of the azimuthal correlation template including the charm feed-down $\lc$ contribution is shown in the left plot of Fig.~\ref{fig:shmrqm} for $3<\ptL<5$ \GeVc and for $0.3<\ptass<1$  \GeVc. Compared to standard PYTHIA 8 predictions with Monash tune, a slightly enhanced production of associated particles in the near-side region is obtained, as most of the charged particles produced from the charm-baryon resonances are collinear to the feed-down $\lc$ baryon. The small amount of the resonance-decay associated tracks produced with larger opening angles with the direction of the trigger $\Lc$ also induces a mild increase of the baseline level, resulting in turn in a decrease of the away-side peak height. The near-side yield trend as a function of $\ptL$ for the same associated particle transverse momentum region is shown in Fig.~\ref{fig:shmrqm} (right). A near-side yield increase between 30\% and 40\% is measured with respect to the standard PYTHIA~8 with Monash tune predictions, in every $\ptL$ interval. Despite this increase, the PYTHIA~8 Monash+Reso predictions are not able to describe the large near-side yield measured at small transverse momentum of the $\lc$. This tends to rule out the presence of additional charm baryon states as the only effect responsible for the large value of the yields observed in data.

%% file: Conclusions.tex
\section{Conclusions}
\label{sec:Conclusions}

The azimuthal correlation distributions between $\Lc$ baryons and charged particles were measured and studied for the first time in pp collisions at $\s = 13$ TeV. The correlation pattern, parametrised in terms of near- and away-side peak associated yields and widths via a fit to the correlation distribution, is sensitive to charm fragmentation and hadronisation. 

In the interval $3 < \ptL < 5$~\GeVc, the associated yield in the near-side peak tends to be higher than that of D mesons correlations with charged particles, albeit with a limited statistical significance of 2.7$\sigma$.
With increasing transverse momentum of the trigger baryon a better agreement is found between the two measurements, pointing towards more similar properties between charm jets containing a $\Lc$ baryon and those containing a D meson. 

The measurement presented in this paper provides interesting insights into the charm-quark fragmentation in baryons through azimuthal correlation distributions, by exposing limitations of the state-of-the-art models that are able to describe the production yield ratios to mesons.
Models that implement different hadronisation mechanisms tend to underestimate the measured near- and away-side peak yields for the low-momentum kinematic regions, although they qualitatively reproduce the yield values for $\ptL >$ 5 \GeVc. 
This consideration also holds for a PYTHIA~8 tune implementing colour-reconnection mechanisms beyond the leading colour approximation, although this model is able to reproduce the $\Lc$-baryon to D-meson production yield ratio within uncertainties. The potential contribution of yet unobserved heavier charm-baryon states, predicted by the RQM model, to the azimuthal correlation distributions, was tested by simulating their decay in PYTHIA~8. The addition of this contribution to the PYTHIA8 predictions does not explain the tension between the model and data at low $\ptL$.  The JETSCAPE model including hybrid hadronisation, where baryon formation is implemented as an interplay of fragmentation and coalescence, provides similar tensions to data as the other models.

The larger data sample and the expected improved performance for tracking and vertexing capabilities of the upgraded ALICE detector in Run 3 will grant significantly smaller uncertainties in future measurements. They could aid in better addressing the properties of the charm fragmentation and hadronisation in jets. This will shed further light on the role of possible different hadronisation processes such as coalescence into charm hadronisation into baryons and on the magnitude of their effect on the correlation patterns and charm-jet shape.

%% file: fa_2024-11-02_Opt_C.tex

The ALICE Collaboration would like to thank all its engineers and technicians for their invaluable contributions to the construction of the experiment and the CERN accelerator teams for the outstanding performance of the LHC complex.
The ALICE Collaboration gratefully acknowledges the resources and support provided by all Grid centres and the Worldwide LHC Computing Grid (WLCG) collaboration.
The ALICE Collaboration acknowledges the following funding agencies for their support in building and running the ALICE detector:
A. I. Alikhanyan National Science Laboratory (Yerevan Physics Institute) Foundation (ANSL), State Committee of Science and World Federation of Scientists (WFS), Armenia;
Austrian Academy of Sciences, Austrian Science Fund (FWF): [M 2467-N36] and Nationalstiftung f\"{u}r Forschung, Technologie und Entwicklung, Austria;
Ministry of Communications and High Technologies, National Nuclear Research Center, Azerbaijan;
Conselho Nacional de Desenvolvimento Cient\'{\i}fico e Tecnol\'{o}gico (CNPq), Financiadora de Estudos e Projetos (Finep), Funda\c{c}\~{a}o de Amparo \`{a} Pesquisa do Estado de S\~{a}o Paulo (FAPESP) and Universidade Federal do Rio Grande do Sul (UFRGS), Brazil;
Bulgarian Ministry of Education and Science, within the National Roadmap for Research Infrastructures 2020-2027 (object CERN), Bulgaria;
Ministry of Education of China (MOEC) , Ministry of Science \& Technology of China (MSTC) and National Natural Science Foundation of China (NSFC), China;
Ministry of Science and Education and Croatian Science Foundation, Croatia;
Centro de Aplicaciones Tecnol\'{o}gicas y Desarrollo Nuclear (CEADEN), Cubaenerg\'{\i}a, Cuba;
Ministry of Education, Youth and Sports of the Czech Republic, Czech Republic;
The Danish Council for Independent Research | Natural Sciences, the VILLUM FONDEN and Danish National Research Foundation (DNRF), Denmark;
Helsinki Institute of Physics (HIP), Finland;
Commissariat \`{a} l'Energie Atomique (CEA) and Institut National de Physique Nucl\'{e}aire et de Physique des Particules (IN2P3) and Centre National de la Recherche Scientifique (CNRS), France;
Bundesministerium f\"{u}r Bildung und Forschung (BMBF) and GSI Helmholtzzentrum f\"{u}r Schwerionenforschung GmbH, Germany;
General Secretariat for Research and Technology, Ministry of Education, Research and Religions, Greece;
National Research, Development and Innovation Office, Hungary;
Department of Atomic Energy Government of India (DAE), Department of Science and Technology, Government of India (DST), University Grants Commission, Government of India (UGC) and Council of Scientific and Industrial Research (CSIR), India;
National Research and Innovation Agency - BRIN, Indonesia;
Istituto Nazionale di Fisica Nucleare (INFN), Italy;
Japanese Ministry of Education, Culture, Sports, Science and Technology (MEXT) and Japan Society for the Promotion of Science (JSPS) KAKENHI, Japan;
Consejo Nacional de Ciencia (CONACYT) y Tecnolog\'{i}a, through Fondo de Cooperaci\'{o}n Internacional en Ciencia y Tecnolog\'{i}a (FONCICYT) and Direcci\'{o}n General de Asuntos del Personal Academico (DGAPA), Mexico;
Nederlandse Organisatie voor Wetenschappelijk Onderzoek (NWO), Netherlands;
The Research Council of Norway, Norway;
Pontificia Universidad Cat\'{o}lica del Per\'{u}, Peru;
Ministry of Science and Higher Education, National Science Centre and WUT ID-UB, Poland;
Korea Institute of Science and Technology Information and National Research Foundation of Korea (NRF), Republic of Korea;
Ministry of Education and Scientific Research, Institute of Atomic Physics, Ministry of Research and Innovation and Institute of Atomic Physics and Universitatea Nationala de Stiinta si Tehnologie Politehnica Bucuresti, Romania;
Ministry of Education, Science, Research and Sport of the Slovak Republic, Slovakia;
National Research Foundation of South Africa, South Africa;
Swedish Research Council (VR) and Knut \& Alice Wallenberg Foundation (KAW), Sweden;
European Organization for Nuclear Research, Switzerland;
Suranaree University of Technology (SUT), National Science and Technology Development Agency (NSTDA) and National Science, Research and Innovation Fund (NSRF via PMU-B B05F650021), Thailand;
Turkish Energy, Nuclear and Mineral Research Agency (TENMAK), Turkey;
National Academy of  Sciences of Ukraine, Ukraine;
Science and Technology Facilities Council (STFC), United Kingdom;
National Science Foundation of the United States of America (NSF) and United States Department of Energy, Office of Nuclear Physics (DOE NP), United States of America.
In addition, individual groups or members have received support from:
Czech Science Foundation (grant no. 23-07499S), Czech Republic;
FORTE project, reg.\ no.\ CZ.02.01.01/00/22\_008/0004632, Czech Republic, co-funded by the European Union, Czech Republic;
European Research Council (grant no. 950692), European Union;
ICSC - Centro Nazionale di Ricerca in High Performance Computing, Big Data and Quantum Computing, European Union - NextGenerationEU;
Academy of Finland (Center of Excellence in Quark Matter) (grant nos. 346327, 346328), Finland;
Deutsche Forschungs Gemeinschaft (DFG, German Research Foundation) ``Neutrinos and Dark Matter in Astro- and Particle Physics'' (grant no. SFB 1258), Germany.

%% file: Alice_Authorlist_2024-11-02_Opt_C.tex
\begin{flushleft} 
\small

S.~Acharya\,\orcidlink{0000-0002-9213-5329}\,$^{\rm 126}$, 
A.~Agarwal$^{\rm 134}$, 
G.~Aglieri Rinella\,\orcidlink{0000-0002-9611-3696}\,$^{\rm 32}$, 
L.~Aglietta\,\orcidlink{0009-0003-0763-6802}\,$^{\rm 24}$, 
M.~Agnello\,\orcidlink{0000-0002-0760-5075}\,$^{\rm 29}$, 
N.~Agrawal\,\orcidlink{0000-0003-0348-9836}\,$^{\rm 25}$, 
Z.~Ahammed\,\orcidlink{0000-0001-5241-7412}\,$^{\rm 134}$, 
S.~Ahmad\,\orcidlink{0000-0003-0497-5705}\,$^{\rm 15}$, 
S.U.~Ahn\,\orcidlink{0000-0001-8847-489X}\,$^{\rm 71}$, 
I.~Ahuja\,\orcidlink{0000-0002-4417-1392}\,$^{\rm 36}$
% ...

\section*{Affiliation Notes}

$^{\rm I}$ Deceased\\
$^{\rm II}$ Also at: Max-Planck-Institut fur Physik, Munich, Germany\\
$^{\rm III}$ Also at: Italian National Agency for New Technologies, Energy and Sustainable Economic Development (ENEA), Bologna, Italy\\
$^{\rm IV}$ Also at: Dipartimento DET del Politecnico di Torino, Turin, Italy\\
$^{\rm V}$ Also at: Department of Applied Physics, Aligarh Muslim University, Aligarh, India\\
$^{\rm VI}$ Also at: Institute of Theoretical Physics, University of Wroclaw, Poland\\
$^{\rm VII}$ Also at: Facultad de Ciencias, Universidad Nacional Autónoma de México, Mexico City, Mexico\\

\section*{Collaboration Institutes}

$^{1}$ A.I. Alikhanyan National Science Laboratory (Yerevan Physics Institute) Foundation, Yerevan, Armenia\\
$^{2}$ AGH University of Krakow, Cracow, Poland\\
$^{3}$ Bogolyubov Institute for Theoretical Physics, National Academy of Sciences of Ukraine, Kiev, Ukraine\\
$^{4}$ Bose Institute, Department of Physics  and Centre for Astroparticle Physics and Space Science (CAPSS), Kolkata, India\\
$^{5}$ California Polytechnic State University, San Luis Obispo, California, United States\\
$^{6}$ Central China Normal University, Wuhan, China\\
$^{7}$ Centro de Aplicaciones Tecnol\'{o}gicas y Desarrollo Nuclear (CEADEN), Havana, Cuba\\
$^{8}$ Centro de Investigaci\'{o}n y de Estudios Avanzados (CINVESTAV), Mexico City and M\'{e}rida, Mexico\\
$^{9}$ Chicago State University, Chicago, Illinois, United States\\
$^{10}$ China Institute of Atomic Energy, Beijing, China\\
$^{11}$ China University of Geosciences, Wuhan, China\\
$^{12}$ Chungbuk National University, Cheongju, Republic of Korea\\
$^{13}$ Comenius University Bratislava, Faculty of Mathematics, Physics and Informatics, Bratislava, Slovak Republic\\
$^{14}$ Creighton University, Omaha, Nebraska, United States\\
$^{15}$ Department of Physics, Aligarh Muslim University, Aligarh, India\\
$^{16}$ Department of Physics, Pusan National University, Pusan, Republic of Korea\\
$^{17}$ Department of Physics, Sejong University, Seoul, Republic of Korea\\
$^{18}$ Department of Physics, University of California, Berkeley, California, United States\\
$^{19}$ Department of Physics, University of Oslo, Oslo, Norway\\
$^{20}$ Department of Physics and Technology, University of Bergen, Bergen, Norway\\
$^{21}$ Dipartimento di Fisica, Universit\`{a} di Pavia, Pavia, Italy\\
$^{22}$ Dipartimento di Fisica dell'Universit\`{a} and Sezione INFN, Cagliari, Italy\\
$^{23}$ Dipartimento di Fisica dell'Universit\`{a} and Sezione INFN, Trieste, Italy\\
$^{24}$ Dipartimento di Fisica dell'Universit\`{a} and Sezione INFN, Turin, Italy\\
$^{25}$ Dipartimento di Fisica e Astronomia dell'Universit\`{a} and Sezione INFN, Bologna, Italy\\
$^{26}$ Dipartimento di Fisica e Astronomia dell'Universit\`{a} and Sezione INFN, Catania, Italy\\
$^{27}$ Dipartimento di Fisica e Astronomia dell'Universit\`{a} and Sezione INFN, Padova, Italy\\
$^{28}$ Dipartimento di Fisica `E.R.~Caianiello' dell'Universit\`{a} and Gruppo Collegato INFN, Salerno, Italy\\
$^{29}$ Dipartimento DISAT del Politecnico and Sezione INFN, Turin, Italy\\
$^{30}$ Dipartimento di Scienze MIFT, Universit\`{a} di Messina, Messina, Italy\\
$^{31}$ Dipartimento Interateneo di Fisica `M.~Merlin' and Sezione INFN, Bari, Italy\\
$^{32}$ European Organization for Nuclear Research (CERN), Geneva, Switzerland\\
$^{33}$ Faculty of Electrical Engineering, Mechanical Engineering and Naval Architecture, University of Split, Split, Croatia\\
$^{34}$ Faculty of Nuclear Sciences and Physical Engineering, Czech Technical University in Prague, Prague, Czech Republic\\
$^{35}$ Faculty of Physics, Sofia University, Sofia, Bulgaria\\
$^{36}$ Faculty of Science, P.J.~\v{S}af\'{a}rik University, Ko\v{s}ice, Slovak Republic\\
$^{37}$ Faculty of Technology, Environmental and Social Sciences, Bergen, Norway\\
$^{38}$ Frankfurt Institute for Advanced Studies, Johann Wolfgang Goethe-Universit\"{a}t Frankfurt, Frankfurt, Germany\\
$^{39}$ Fudan University, Shanghai, China\\
$^{40}$ Gangneung-Wonju National University, Gangneung, Republic of Korea\\
$^{41}$ Gauhati University, Department of Physics, Guwahati, India\\
$^{42}$ Helmholtz-Institut f\"{u}r Strahlen- und Kernphysik, Rheinische Friedrich-Wilhelms-Universit\"{a}t Bonn, Bonn, Germany\\
$^{43}$ Helsinki Institute of Physics (HIP), Helsinki, Finland\\
$^{44}$ High Energy Physics Group,  Universidad Aut\'{o}noma de Puebla, Puebla, Mexico\\
$^{45}$ Horia Hulubei National Institute of Physics and Nuclear Engineering, Bucharest, Romania\\
$^{46}$ HUN-REN Wigner Research Centre for Physics, Budapest, Hungary\\
$^{47}$ Indian Institute of Technology Bombay (IIT), Mumbai, India\\
$^{48}$ Indian Institute of Technology Indore, Indore, India\\
$^{49}$ INFN, Laboratori Nazionali di Frascati, Frascati, Italy\\
$^{50}$ INFN, Sezione di Bari, Bari, Italy\\
$^{51}$ INFN, Sezione di Bologna, Bologna, Italy\\
$^{52}$ INFN, Sezione di Cagliari, Cagliari, Italy\\
$^{53}$ INFN, Sezione di Catania, Catania, Italy\\
$^{54}$ INFN, Sezione di Padova, Padova, Italy\\
$^{55}$ INFN, Sezione di Pavia, Pavia, Italy\\
$^{56}$ INFN, Sezione di Torino, Turin, Italy\\
$^{57}$ INFN, Sezione di Trieste, Trieste, Italy\\
$^{58}$ Inha University, Incheon, Republic of Korea\\
$^{59}$ Institute for Gravitational and Subatomic Physics (GRASP), Utrecht University/Nikhef, Utrecht, Netherlands\\
$^{60}$ Institute of Experimental Physics, Slovak Academy of Sciences, Ko\v{s}ice, Slovak Republic\\
$^{61}$ Institute of Physics, Homi Bhabha National Institute, Bhubaneswar, India\\
$^{62}$ Institute of Physics of the Czech Academy of Sciences, Prague, Czech Republic\\
$^{63}$ Institute of Space Science (ISS), Bucharest, Romania\\
$^{64}$ Institut f\"{u}r Kernphysik, Johann Wolfgang Goethe-Universit\"{a}t Frankfurt, Frankfurt, Germany\\
$^{65}$ Instituto de Ciencias Nucleares, Universidad Nacional Aut\'{o}noma de M\'{e}xico, Mexico City, Mexico\\
$^{66}$ Instituto de F\'{i}sica, Universidade Federal do Rio Grande do Sul (UFRGS), Porto Alegre, Brazil\\
$^{67}$ Instituto de F\'{\i}sica, Universidad Nacional Aut\'{o}noma de M\'{e}xico, Mexico City, Mexico\\
$^{68}$ iThemba LABS, National Research Foundation, Somerset West, South Africa\\
$^{69}$ Jeonbuk National University, Jeonju, Republic of Korea\\
$^{70}$ Johann-Wolfgang-Goethe Universit\"{a}t Frankfurt Institut f\"{u}r Informatik, Fachbereich Informatik und Mathematik, Frankfurt, Germany\\
$^{71}$ Korea Institute of Science and Technology Information, Daejeon, Republic of Korea\\
$^{72}$ Laboratoire de Physique Subatomique et de Cosmologie, Universit\'{e} Grenoble-Alpes, CNRS-IN2P3, Grenoble, France\\
$^{73}$ Lawrence Berkeley National Laboratory, Berkeley, California, United States\\
$^{74}$ Lund University Department of Physics, Division of Particle Physics, Lund, Sweden\\
$^{75}$ Nagasaki Institute of Applied Science, Nagasaki, Japan\\
$^{76}$ Nara Women{'}s University (NWU), Nara, Japan\\
$^{77}$ National and Kapodistrian University of Athens, School of Science, Department of Physics , Athens, Greece\\
$^{78}$ National Centre for Nuclear Research, Warsaw, Poland\\
$^{79}$ National Institute of Science Education and Research, Homi Bhabha National Institute, Jatni, India\\
$^{80}$ National Nuclear Research Center, Baku, Azerbaijan\\
$^{81}$ National Research and Innovation Agency - BRIN, Jakarta, Indonesia\\
$^{82}$ Niels Bohr Institute, University of Copenhagen, Copenhagen, Denmark\\
$^{83}$ Nikhef, National institute for subatomic physics, Amsterdam, Netherlands\\
$^{84}$ Nuclear Physics Group, STFC Daresbury Laboratory, Daresbury, United Kingdom\\
$^{85}$ Nuclear Physics Institute of the Czech Academy of Sciences, Husinec-\v{R}e\v{z}, Czech Republic\\
$^{86}$ Oak Ridge National Laboratory, Oak Ridge, Tennessee, United States\\
$^{87}$ Ohio State University, Columbus, Ohio, United States\\
$^{88}$ Physics department, Faculty of science, University of Zagreb, Zagreb, Croatia\\
$^{89}$ Physics Department, Panjab University, Chandigarh, India\\
$^{90}$ Physics Department, University of Jammu, Jammu, India\\
$^{91}$ Physics Program and International Institute for Sustainability with Knotted Chiral Meta Matter (WPI-SKCM$^{2}$), Hiroshima University, Hiroshima, Japan\\
$^{92}$ Physikalisches Institut, Eberhard-Karls-Universit\"{a}t T\"{u}bingen, T\"{u}bingen, Germany\\
$^{93}$ Physikalisches Institut, Ruprecht-Karls-Universit\"{a}t Heidelberg, Heidelberg, Germany\\
$^{94}$ Physik Department, Technische Universit\"{a}t M\"{u}nchen, Munich, Germany\\
$^{95}$ Politecnico di Bari and Sezione INFN, Bari, Italy\\
$^{96}$ Research Division and ExtreMe Matter Institute EMMI, GSI Helmholtzzentrum f\"ur Schwerionenforschung GmbH, Darmstadt, Germany\\
$^{97}$ Saga University, Saga, Japan\\
$^{98}$ Saha Institute of Nuclear Physics, Homi Bhabha National Institute, Kolkata, India\\
$^{99}$ School of Physics and Astronomy, University of Birmingham, Birmingham, United Kingdom\\
$^{100}$ Secci\'{o}n F\'{\i}sica, Departamento de Ciencias, Pontificia Universidad Cat\'{o}lica del Per\'{u}, Lima, Peru\\
$^{101}$ Stefan Meyer Institut f\"{u}r Subatomare Physik (SMI), Vienna, Austria\\
$^{102}$ SUBATECH, IMT Atlantique, Nantes Universit\'{e}, CNRS-IN2P3, Nantes, France\\
$^{103}$ Sungkyunkwan University, Suwon City, Republic of Korea\\
$^{104}$ Suranaree University of Technology, Nakhon Ratchasima, Thailand\\
$^{105}$ Technical University of Ko\v{s}ice, Ko\v{s}ice, Slovak Republic\\
$^{106}$ The Henryk Niewodniczanski Institute of Nuclear Physics, Polish Academy of Sciences, Cracow, Poland\\
$^{107}$ The University of Texas at Austin, Austin, Texas, United States\\
$^{108}$ Universidad Aut\'{o}noma de Sinaloa, Culiac\'{a}n, Mexico\\
$^{109}$ Universidade de S\~{a}o Paulo (USP), S\~{a}o Paulo, Brazil\\
$^{110}$ Universidade Estadual de Campinas (UNICAMP), Campinas, Brazil\\
$^{111}$ Universidade Federal do ABC, Santo Andre, Brazil\\
$^{112}$ Universitatea Nationala de Stiinta si Tehnologie Politehnica Bucuresti, Bucharest, Romania\\
$^{113}$ University of Cape Town, Cape Town, South Africa\\
$^{114}$ University of Derby, Derby, United Kingdom\\
$^{115}$ University of Houston, Houston, Texas, United States\\
$^{116}$ University of Jyv\"{a}skyl\"{a}, Jyv\"{a}skyl\"{a}, Finland\\
$^{117}$ University of Kansas, Lawrence, Kansas, United States\\
$^{118}$ University of Liverpool, Liverpool, United Kingdom\\
$^{119}$ University of Science and Technology of China, Hefei, China\\
$^{120}$ University of South-Eastern Norway, Kongsberg, Norway\\
$^{121}$ University of Tennessee, Knoxville, Tennessee, United States\\
$^{122}$ University of the Witwatersrand, Johannesburg, South Africa\\
$^{123}$ University of Tokyo, Tokyo, Japan\\
$^{124}$ University of Tsukuba, Tsukuba, Japan\\
$^{125}$ Universit\"{a}t M\"{u}nster, Institut f\"{u}r Kernphysik, M\"{u}nster, Germany\\
$^{126}$ Universit\'{e} Clermont Auvergne, CNRS/IN2P3, LPC, Clermont-Ferrand, France\\
$^{127}$ Universit\'{e} de Lyon, CNRS/IN2P3, Institut de Physique des 2 Infinis de Lyon, Lyon, France\\
$^{128}$ Universit\'{e} de Strasbourg, CNRS, IPHC UMR 7178, F-67000 Strasbourg, France, Strasbourg, France\\
$^{129}$ Universit\'{e} Paris-Saclay, Centre d'Etudes de Saclay (CEA), IRFU, D\'{e}partment de Physique Nucl\'{e}aire (DPhN), Saclay, France\\
$^{130}$ Universit\'{e}  Paris-Saclay, CNRS/IN2P3, IJCLab, Orsay, France\\
$^{131}$ Universit\`{a} degli Studi di Foggia, Foggia, Italy\\
$^{132}$ Universit\`{a} del Piemonte Orientale, Vercelli, Italy\\
$^{133}$ Universit\`{a} di Brescia, Brescia, Italy\\
$^{134}$ Variable Energy Cyclotron Centre, Homi Bhabha National Institute, Kolkata, India\\
$^{135}$ Warsaw University of Technology, Warsaw, Poland\\
$^{136}$ Wayne State University, Detroit, Michigan, United States\\
$^{137}$ Yale University, New Haven, Connecticut, United States\\
$^{138}$ Yildiz Technical University, Istanbul, Turkey\\
$^{139}$ Yonsei University, Seoul, Republic of Korea\\
$^{140}$ Affiliated with an institute covered by a cooperation agreement with CERN\\
$^{141}$ Affiliated with an international laboratory covered by a cooperation agreement with CERN.\\

\end{flushleft} 

%% file: main.bbl
\providecommand{\href}[2]{#2}\begingroup\raggedright\begin{thebibliography}{10}

\bibitem{Collins:1989gx}
J.~C. Collins, D.~E. Soper, and G.~F. Sterman, ``{Factorization of Hard
  Processes in QCD}'', \href{https://doi.org/10.1142/9789814503266_0001}{{\em
  Adv. Ser. Direct. High Energy Phys.} {\bfseries 5} (1989) 1--91},
  \href{https://arxiv.org/abs/hep-ph/0409313}{{\ttfamily
  arXiv:hep-ph/0409313}}.

\bibitem{ParticleDataGroup:2024cfk}
{\bfseries Particle Data Group} Collaboration, S.~Navas {\em et~al.}, ``{Review
  of particle physics}'',
  \href{https://doi.org/10.1103/PhysRevD.110.030001}{{\em Phys. Rev. D}
  {\bfseries 110} (2024) 030001}.

\bibitem{ALICE:2019nxm}
{\bfseries ALICE} Collaboration, S.~Acharya {\em et~al.}, ``{Measurement of
  ${{\mathrm{D}}^0}$ , ${{\mathrm{D}}^+}$ , ${{\mathrm{D}}^{*+}}$ and
  ${{\mathrm{D}}^+_{\mathrm{s}}}$ production in pp collisions at
  ${\sqrt{{\textit{s}}}~=~5.02~{\text {TeV}}}$ with ALICE}'',
  \href{https://doi.org/10.1140/epjc/s10052-019-6873-6}{{\em Eur. Phys. J. C}
  {\bfseries 79} (2019) 388},
  \href{https://arxiv.org/abs/1901.07979}{{\ttfamily arXiv:1901.07979
  [nucl-ex]}}.

\bibitem{ALICE:2021npz}
{\bfseries ALICE} Collaboration, S.~Acharya {\em et~al.}, ``{Observation of a
  multiplicity dependence in the $\pt$-differential charm baryon-to-meson
  ratios in proton\textendash{}proton collisions at $\sqrt{s}$=13 TeV}'',
  \href{https://doi.org/10.1016/j.physletb.2022.137065}{{\em Phys. Lett. B}
  {\bfseries 829} (2022) 137065},
  \href{https://arxiv.org/abs/2111.11948}{{\ttfamily arXiv:2111.11948
  [nucl-ex]}}.

\bibitem{ALICE:2023sgl}
{\bfseries ALICE} Collaboration, S.~Acharya {\em et~al.}, ``{Charm production
  and fragmentation fractions at midrapidity in pp collisions at $
  \sqrt{\textrm{s}} $ = 13 TeV}'',
  \href{https://doi.org/10.1007/JHEP12(2023)086}{{\em JHEP} {\bfseries 12}
  (2023) 086}, \href{https://arxiv.org/abs/2308.04877}{{\ttfamily
  arXiv:2308.04877 [hep-ex]}}.

\bibitem{ATLAS:2015igt}
{\bfseries ATLAS} Collaboration, G.~Aad {\em et~al.}, ``{Measurement of
  $\mathrm{D}^{*\pm}$, $\mathrm{D}^\pm$ and $\mathrm{D_s}^\pm$ meson production
  cross sections in pp collisions at $\sqrt{s}=7$ TeV with the ATLAS
  detector}'', \href{https://doi.org/10.1016/j.nuclphysb.2016.04.032}{{\em
  Nucl. Phys. B} {\bfseries 907} (2016) 717--763},
  \href{https://arxiv.org/abs/1512.02913}{{\ttfamily arXiv:1512.02913
  [hep-ex]}}.

\bibitem{CMS:2021lab}
{\bfseries CMS} Collaboration, A.~Tumasyan {\em et~al.}, ``{Measurement of
  prompt open-charm production cross sections in proton-proton collisions at $
  \sqrt{s} $ = 13 TeV}'', \href{https://doi.org/10.1007/JHEP11(2021)225}{{\em
  JHEP} {\bfseries 11} (2021) 225},
  \href{https://arxiv.org/abs/2107.01476}{{\ttfamily arXiv:2107.01476
  [hep-ex]}}.

\bibitem{LHCb:2016ikn}
{\bfseries LHCb} Collaboration, R.~Aaij {\em et~al.}, ``{Measurements of prompt
  charm production cross-sections in pp collisions at $ \sqrt{s}=5 $ TeV}'',
  \href{https://doi.org/10.1007/JHEP06(2017)147}{{\em JHEP} {\bfseries 06}
  (2017) 147}, \href{https://arxiv.org/abs/1610.02230}{{\ttfamily
  arXiv:1610.02230 [hep-ex]}}.

\bibitem{LHCb:2015swx}
{\bfseries LHCb} Collaboration, R.~Aaij {\em et~al.}, ``{Measurements of prompt
  charm production cross-sections in pp collisions at $ \sqrt{s}=13 $ TeV}'',
  \href{https://doi.org/10.1007/JHEP03(2016)159}{{\em JHEP} {\bfseries 03}
  (2016) 159}, \href{https://arxiv.org/abs/1510.01707}{{\ttfamily
  arXiv:1510.01707 [hep-ex]}}. [Erratum: JHEP 09, 013 (2016), Erratum: JHEP 05,
  074 (2017)].

\bibitem{ALICE:2022exq}
{\bfseries ALICE} Collaboration, S.~Acharya {\em et~al.}, ``{First measurement
  of $\Lambda_\mathrm{c}^{+}$ production down to $\pt$=0 in pp and p--Pb
  collisions at $\sqrt{s_\mathrm{NN}}$=5.02 TeV}'',
  \href{https://doi.org/10.1103/PhysRevC.107.064901}{{\em Phys. Rev. C}
  {\bfseries 107} (2023) 064901},
  \href{https://arxiv.org/abs/2211.14032}{{\ttfamily arXiv:2211.14032
  [nucl-ex]}}.

\bibitem{ALICE:2020wla}
{\bfseries ALICE} Collaboration, S.~Acharya {\em et~al.},
  ``{$\Lambda_\mathrm{c}^{+}$ production in pp and in p--Pb collisions at
  $\sqrt {s_\mathrm{NN}}$=5.02 TeV}'',
  \href{https://doi.org/10.1103/PhysRevC.104.054905}{{\em Phys. Rev. C}
  {\bfseries 104} (2021) 054905},
  \href{https://arxiv.org/abs/2011.06079}{{\ttfamily arXiv:2011.06079
  [nucl-ex]}}.

\bibitem{ALICE:2020wfu}
{\bfseries ALICE} Collaboration, S.~Acharya {\em et~al.}, ``{$\Lambda^+_c$
  Production and Baryon-to-Meson Ratios in pp and p-Pb Collisions at $\sqrt
  {s_\mathrm{NN}}$=5.02\,\,TeV at the LHC}'',
  \href{https://doi.org/10.1103/PhysRevLett.127.202301}{{\em Phys. Rev. Lett.}
  {\bfseries 127} (2021) 202301},
  \href{https://arxiv.org/abs/2011.06078}{{\ttfamily arXiv:2011.06078
  [nucl-ex]}}.

\bibitem{ALICE:2021rzj}
{\bfseries ALICE} Collaboration, S.~Acharya {\em et~al.}, ``{Measurement of
  Prompt D$^{0}$, $\Lambda_{c}^{+}$, and $\Sigma_{c}^{0,++}$(2455) Production
  in Proton\textendash{}Proton Collisions at $\sqrt s$ = 13\,\,TeV}'',
  \href{https://doi.org/10.1103/PhysRevLett.128.012001}{{\em Phys. Rev. Lett.}
  {\bfseries 128} (2022) 012001},
  \href{https://arxiv.org/abs/2106.08278}{{\ttfamily arXiv:2106.08278
  [hep-ex]}}.

\bibitem{CMS:2019uws}
{\bfseries CMS} Collaboration, A.~M. Sirunyan {\em et~al.}, ``{Production of
  $\Lambda_\mathrm{c}^+$ baryons in proton-proton and lead-lead collisions at
  $\sqrt{s_\mathrm{NN}}=$ 5.02 TeV}'',
  \href{https://doi.org/10.1016/j.physletb.2020.135328}{{\em Phys. Lett. B}
  {\bfseries 803} (2020) 135328},
  \href{https://arxiv.org/abs/1906.03322}{{\ttfamily arXiv:1906.03322
  [hep-ex]}}.

\bibitem{ALICE:2021psx}
{\bfseries ALICE} Collaboration, S.~Acharya {\em et~al.}, ``{Measurement of the
  production cross section of prompt $ {\Xi}_{\mathrm{c}}^0 $ baryons at
  midrapidity in pp collisions at $ \sqrt{s} $ = 5.02 TeV}'',
  \href{https://doi.org/10.1007/JHEP10(2021)159}{{\em JHEP} {\bfseries 10}
  (2021) 159}, \href{https://arxiv.org/abs/2105.05616}{{\ttfamily
  arXiv:2105.05616 [nucl-ex]}}.

\bibitem{ALICE:2021bli}
{\bfseries ALICE} Collaboration, S.~Acharya {\em et~al.}, ``{Measurement of the
  Cross Sections of $\Xi^0_{c}$ and $\Xi^+_{c}$ Baryons and of the
  Branching-Fraction Ratio BR($\Xi^0_\mathrm{c} \rightarrow \Xi^-{e}^+\nu_{
  e}$)/BR($\Xi^0_\mathrm{c} \rightarrow \Xi^-\pi^+$) in pp collisions at 13
  TeV}'', \href{https://doi.org/10.1103/PhysRevLett.127.272001}{{\em Phys. Rev.
  Lett.} {\bfseries 127} (2021) 272001},
  \href{https://arxiv.org/abs/2105.05187}{{\ttfamily arXiv:2105.05187
  [nucl-ex]}}.

\bibitem{ALICE:2022cop}
{\bfseries ALICE} Collaboration, S.~Acharya {\em et~al.}, ``{First measurement
  of $\Omega_\mathrm{c}^{0}$ production in pp collisions at $\sqrt{s}$=13
  TeV}'', \href{https://doi.org/10.1016/j.physletb.2022.137625}{{\em Phys.
  Lett. B} {\bfseries 846} (2023) 137625},
  \href{https://arxiv.org/abs/2205.13993}{{\ttfamily arXiv:2205.13993
  [nucl-ex]}}.

\bibitem{ALICE:2021dhb}
{\bfseries ALICE} Collaboration, S.~Acharya {\em et~al.}, ``{Charm-quark
  fragmentation fractions and production cross section at midrapidity in pp
  collisions at the LHC}'',
  \href{https://doi.org/10.1103/PhysRevD.105.L011103}{{\em Phys. Rev. D}
  {\bfseries 105} (2022) L011103},
  \href{https://arxiv.org/abs/2105.06335}{{\ttfamily arXiv:2105.06335
  [nucl-ex]}}.

\bibitem{Lisovyi:2015uqa}
M.~Lisovyi, A.~Verbytskyi, and O.~Zenaiev, ``{Combined analysis of charm-quark
  fragmentation-fraction measurements}'',
  \href{https://doi.org/10.1140/epjc/s10052-016-4246-y}{{\em Eur. Phys. J. C}
  {\bfseries 76} (2016) 397},
  \href{https://arxiv.org/abs/1509.01061}{{\ttfamily arXiv:1509.01061
  [hep-ex]}}.

\bibitem{Azizi:2014nta}
K.~Azizi, H.~Sundu, and S.~Sahin, ``{Semileptonic $B_\mathrm{s} \rightarrow
  D_\mathrm{s2}^{*}(2573)\ell\bar{\nu}_{\ell}$ transition in QCD}'',
  \href{https://doi.org/10.1140/epjc/s10052-015-3402-0}{{\em Eur. Phys. J. C}
  {\bfseries 75} (2015) 197}, \href{https://arxiv.org/abs/1411.3100}{{\ttfamily
  arXiv:1411.3100 [hep-ph]}}.

\bibitem{Christiansen:2015yqa}
J.~R. Christiansen and P.~Z. Skands, ``{String Formation Beyond Leading
  Colour}'', \href{https://doi.org/10.1007/JHEP08(2015)003}{{\em JHEP}
  {\bfseries 08} (2015) 003},
  \href{https://arxiv.org/abs/1505.01681}{{\ttfamily arXiv:1505.01681
  [hep-ph]}}.

\bibitem{Song:2018tpv}
J.~Song, H.-h. Li, and F.-l. Shao, ``{New feature of low $p_\mathrm{T}$ charm
  quark hadronization in pp collisions at $\sqrt{s}=7$ TeV}'',
  \href{https://doi.org/10.1140/epjc/s10052-018-5817-x}{{\em Eur. Phys. J. C}
  {\bfseries 78} (2018) 344},
  \href{https://arxiv.org/abs/1801.09402}{{\ttfamily arXiv:1801.09402
  [hep-ph]}}.

\bibitem{Minissale:2020bif}
V.~Minissale, S.~Plumari, and V.~Greco, ``{Charm hadrons in pp collisions at
  LHC energy within a coalescence plus fragmentation approach}'',
  \href{https://doi.org/10.1016/j.physletb.2021.136622}{{\em Phys. Lett. B}
  {\bfseries 821} (2021) 136622},
  \href{https://arxiv.org/abs/2012.12001}{{\ttfamily arXiv:2012.12001
  [hep-ph]}}.

\bibitem{He:2019tik}
M.~He and R.~Rapp, ``{Charm-Baryon Production in Proton-Proton Collisions}'',
  \href{https://doi.org/10.1016/j.physletb.2019.06.004}{{\em Phys. Lett. B}
  {\bfseries 795} (2019) 117--121},
  \href{https://arxiv.org/abs/1902.08889}{{\ttfamily arXiv:1902.08889
  [nucl-th]}}.

\bibitem{ALICE:2023jgm}
{\bfseries ALICE} Collaboration, S.~Acharya {\em et~al.}, ``{Exploring the
  non-universality of charm hadronisation through the measurement of the
  fraction of jet longitudinal momentum carried by $\Lambda_{\rm c}^+$ baryons
  in pp collisions}'', \href{https://arxiv.org/abs/2301.13798}{{\ttfamily
  arXiv:2301.13798 [nucl-ex]}}.

\bibitem{ALICE:2022mur}
{\bfseries ALICE} Collaboration, S.~Acharya {\em et~al.}, ``{Measurement of the
  production of charm jets tagged with D$^{0}$ mesons in pp collisions at $
  \sqrt{s} $ = 5.02 and 13 TeV}'',
  \href{https://doi.org/10.1007/JHEP06(2023)133}{{\em JHEP} {\bfseries 06}
  (2023) 133}, \href{https://arxiv.org/abs/2204.10167}{{\ttfamily
  arXiv:2204.10167 [nucl-ex]}}.

\bibitem{ALICE:2021kpy}
{\bfseries ALICE} Collaboration, S.~Acharya {\em et~al.}, ``{Investigating
  charm production and fragmentation via azimuthal correlations of prompt D
  mesons with charged particles in pp collisions at $\sqrt{s} = 13$ TeV}'',
  \href{https://doi.org/10.1140/epjc/s10052-022-10267-3}{{\em Eur. Phys. J. C}
  {\bfseries 82} (2022) 335},
  \href{https://arxiv.org/abs/2110.10043}{{\ttfamily arXiv:2110.10043
  [nucl-ex]}}.

\bibitem{ALICE:2016clc}
{\bfseries ALICE} Collaboration, J.~Adam {\em et~al.}, ``{Measurement of
  azimuthal correlations of D mesons and charged particles in pp collisions at
  $\sqrt{s}=7$ TeV and p--Pb collisions at $\sqrt{s_{\rm NN}}=5.02$ TeV}'',
  \href{https://doi.org/10.1140/epjc/s10052-017-4779-8}{{\em Eur. Phys. J. C}
  {\bfseries 77} (2017) 245},
  \href{https://arxiv.org/abs/1605.06963}{{\ttfamily arXiv:1605.06963
  [nucl-ex]}}.

\bibitem{ALICE:2019oyn}
{\bfseries ALICE} Collaboration, S.~Acharya {\em et~al.}, ``{Azimuthal
  correlations of prompt D mesons with charged particles in pp and
  p\textendash{}Pb collisions at $\sqrt{s_{NN}}$ = 5.02 TeV}'',
  \href{https://doi.org/10.1140/epjc/s10052-020-8118-0}{{\em Eur. Phys. J. C}
  {\bfseries 80} (2020) 979},
  \href{https://arxiv.org/abs/1910.14403}{{\ttfamily arXiv:1910.14403
  [nucl-ex]}}.

\bibitem{Aamodt:2008zz}
{\bfseries ALICE} Collaboration, K.~Aamodt {\em et~al.}, ``{The ALICE
  experiment at the CERN LHC}'',
  \href{https://doi.org/10.1088/1748-0221/3/08/S08002}{{\em JINST} {\bfseries
  3} (2008) S08002}.

\bibitem{Abelev:2014ffa}
{\bfseries ALICE} Collaboration, B.~Abelev {\em et~al.}, ``{Performance of the
  ALICE Experiment at the CERN LHC}'',
  \href{https://doi.org/10.1142/S0217751X14300440}{{\em Int. J. Mod. Phys. A}
  {\bfseries 29} (2014) 1430044},
  \href{https://arxiv.org/abs/1402.4476}{{\ttfamily arXiv:1402.4476
  [nucl-ex]}}.

\bibitem{Aamodt:2010aa}
{\bfseries ALICE} Collaboration, K.~Aamodt {\em et~al.}, ``{Alignment of the
  ALICE Inner Tracking System with cosmic-ray tracks}'',
  \href{https://doi.org/10.1088/1748-0221/5/03/P03003}{{\em JINST} {\bfseries
  5} (2010) P03003}, \href{https://arxiv.org/abs/1001.0502}{{\ttfamily
  arXiv:1001.0502 [physics.ins-det]}}.

\bibitem{Alme:2010ke}
J.~Alme {\em et~al.}, ``{The ALICE TPC, a large 3-dimensional tracking device
  with fast readout for ultra-high multiplicity events}'',
  \href{https://doi.org/10.1016/j.nima.2010.04.042}{{\em Nucl. Instrum. Meth.
  A} {\bfseries 622} (2010) 316--367},
  \href{https://arxiv.org/abs/1001.1950}{{\ttfamily arXiv:1001.1950
  [physics.ins-det]}}.

\bibitem{Akindinov:2013tea}
A.~Akindinov {\em et~al.}, ``{Performance of the ALICE Time-Of-Flight detector
  at the LHC}'', \href{https://doi.org/10.1140/epjp/i2013-13044-x}{{\em Eur.
  Phys. J. Plus} {\bfseries 128} (2013) 44}.

\bibitem{Abbas:2013taa}
{\bfseries ALICE} Collaboration, E.~Abbas {\em et~al.}, ``{Performance of the
  ALICE VZERO system}'',
  \href{https://doi.org/10.1088/1748-0221/8/10/P10016}{{\em JINST} {\bfseries
  8} (2013) P10016},
\href{https://arxiv.org/abs/1306.3130}{{\ttfamily arXiv:1306.3130 [nucl-ex]}}.

\bibitem{Adam:2016ilk}
{\bfseries ALICE} Collaboration, J.~Adam {\em et~al.}, ``{Determination of the
  event collision time with the ALICE detector at the LHC}'',
  \href{https://doi.org/10.1140/epjp/i2017-11279-1}{{\em Eur. Phys. J. Plus}
  {\bfseries 132} (2017) 99},
\href{https://arxiv.org/abs/1610.03055}{{\ttfamily arXiv:1610.03055
  [physics.ins-det]}}.

\bibitem{ALICE-PUBLIC-2016-002}
{\bfseries ALICE} Collaboration, S.~Acharya {\em et~al.}, ``{ALICE
  2016-2017-2018 luminosity determination for pp collisions at $\sqrt{s}$ = 13
  TeV}.'' Jul, 2021.
\newblock \url{https://cds.cern.ch/record/2776672}. ALICE-PUBLIC-2021-005.

\bibitem{Sjostrand:2014zea}
T.~Sj\"ostrand, S.~Ask, J.~R. Christiansen, R.~Corke, N.~Desai, P.~Ilten,
  S.~Mrenna, S.~Prestel, C.~O. Rasmussen, and P.~Z. Skands, ``{An introduction
  to PYTHIA 8.2}'', \href{https://doi.org/10.1016/j.cpc.2015.01.024}{{\em
  Comput. Phys. Commun.} {\bfseries 191} (2015) 159--177},
  \href{https://arxiv.org/abs/1410.3012}{{\ttfamily arXiv:1410.3012 [hep-ph]}}.

\bibitem{Skands:2014pea}
P.~Skands, S.~Carrazza, and J.~Rojo, ``{Tuning PYTHIA 8.1: the Monash 2013
  Tune}'', \href{https://doi.org/10.1140/epjc/s10052-014-3024-y}{{\em Eur.
  Phys. J. C} {\bfseries 74} (2014) 3024},
  \href{https://arxiv.org/abs/1404.5630}{{\ttfamily arXiv:1404.5630 [hep-ph]}}.

\bibitem{Brun:1082634}
R.~Brun, F.~Bruyant, F.~Carminati, S.~Giani, M.~Maire, A.~McPherson,
  G.~Patrick, and L.~Urban, \href{https://doi.org/10.17181/CERN.MUHF.DMJ1}{{\em
  {GEANT: Detector Description and Simulation Tool; Oct 1994}}}.
\newblock CERN Program Library. CERN, Geneva, 1993.
\newblock \url{http://cds.cern.ch/record/1082634}.
\newblock Long Writeup W5013.

\bibitem{2016arXiv160302754C}
T.~Chen and C.~Guestrin,
  \href{https://doi.org/10.1145/2939672.2939785}{``{XGBoost: A Scalable Tree
  Boosting System}'',} in {\em Proceedings of the 22nd ACM SIGKDD International
  Conference on Knowledge Discovery and Data Mining}, KDD~'16, pp.~785--794.
\newblock ACM, Aug., 2016.
\newblock \href{https://arxiv.org/abs/1603.02754}{{\ttfamily arXiv:1603.02754
  [cs.LG]}}.

\bibitem{ALICE-PUBLIC-2017-005}
{\bfseries ALICE} Collaboration, S.~Acharya {\em et~al.}, ``The alice
  definition of primary particles.'' Jun, 2017.
\newblock \url{https://cds.cern.ch/record/2270008}. ALICE-PUBLIC-2017-005.

\bibitem{Cacciari:1998it}
M.~Cacciari, M.~Greco, and P.~Nason, ``{The $p_\mathrm{T}$ spectrum in
  heavy-flavour hadroproduction.}'',
  \href{https://doi.org/10.1088/1126-6708/1998/05/007}{{\em JHEP} {\bfseries
  05} (1998) 007}, \href{https://arxiv.org/abs/hep-ph/9803400}{{\ttfamily
  arXiv:hep-ph/9803400}}.

\bibitem{Cacciari:2001td}
M.~Cacciari, S.~Frixione, and P.~Nason, ``{The $\pt$ spectrum in heavy flavor
  photoproduction}'', \href{https://doi.org/10.1088/1126-6708/2001/03/006}{{\em
  JHEP} {\bfseries 03} (2001) 006},
  \href{https://arxiv.org/abs/hep-ph/0102134}{{\ttfamily
  arXiv:hep-ph/0102134}}.

\bibitem{Cacciari:2012ny}
M.~Cacciari, S.~Frixione, N.~Houdeau, M.~L. Mangano, P.~Nason, and G.~Ridolfi,
  ``{Theoretical predictions for charm and bottom production at the LHC}'',
  \href{https://doi.org/10.1007/JHEP10(2012)137}{{\em JHEP} {\bfseries 10}
  (2012) 137}, \href{https://arxiv.org/abs/1205.6344}{{\ttfamily
  arXiv:1205.6344 [hep-ph]}}.

\bibitem{LHCb:2019fns}
{\bfseries LHCb} Collaboration, R.~Aaij {\em et~al.}, ``{Measurement of $b$
  hadron fractions in 13 TeV $pp$ collisions}'',
  \href{https://doi.org/10.1103/PhysRevD.100.031102}{{\em Phys. Rev. D}
  {\bfseries 100} (2019) 031102},
  \href{https://arxiv.org/abs/1902.06794}{{\ttfamily arXiv:1902.06794
  [hep-ex]}}.

\bibitem{upton2008dictionary}
G.~Upton and I.~Cook, {\em A Dictionary of Statistics}.
\newblock Oxford Paperback Reference. OUP Oxford, 2008.
\newblock \url{https://doi.org/10.1093/acref/9780199679188.001.0001}.

\bibitem{Nason:2004rx}
P.~Nason, ``{A New method for combining NLO QCD with shower Monte Carlo
  algorithms}'', \href{https://doi.org/10.1088/1126-6708/2004/11/040}{{\em
  JHEP} {\bfseries 11} (2004) 040},
  \href{https://arxiv.org/abs/hep-ph/0409146}{{\ttfamily
  arXiv:hep-ph/0409146}}.

\bibitem{Frixione:2007vw}
S.~Frixione, P.~Nason, and C.~Oleari, ``{Matching NLO QCD computations with
  Parton Shower simulations: the POWHEG method}'',
  \href{https://doi.org/10.1088/1126-6708/2007/11/070}{{\em JHEP} {\bfseries
  11} (2007) 070}, \href{https://arxiv.org/abs/0709.2092}{{\ttfamily
  arXiv:0709.2092 [hep-ph]}}.

\bibitem{JETSCAPE:2019udz}
{\bfseries JETSCAPE} Collaboration, A.~Kumar {\em et~al.}, ``{JETSCAPE
  framework: $p+p$ results}'',
  \href{https://doi.org/10.1103/PhysRevC.102.054906}{{\em Phys. Rev. C}
  {\bfseries 102} (2020) 054906},
  \href{https://arxiv.org/abs/1910.05481}{{\ttfamily arXiv:1910.05481
  [nucl-th]}}.

\bibitem{Han:2016uhh}
K.~C. Han, R.~J. Fries, and C.~M. Ko, ``{Jet Fragmentation via Recombination of
  Parton Showers}'', \href{https://doi.org/10.1103/PhysRevC.93.045207}{{\em
  Phys. Rev. C} {\bfseries 93} (2016) 045207},
  \href{https://arxiv.org/abs/1601.00708}{{\ttfamily arXiv:1601.00708
  [nucl-th]}}.

\end{thebibliography}\endgroup
